\expandafter\def\csname ver@fixltx2e.sty\endcsname{} 
\documentclass[journal]{IEEEtran}
\usepackage{cite}
\usepackage{float}
\usepackage{graphicx} 
\usepackage{tikz}
\usepackage{pgfplots}
\pgfplotsset{compat=1.15}
\usepackage{bm}
\usepackage{color}
\usepackage{xcolor}
\usepackage{amstext,epsfig,amssymb,amsbsy,amsmath}
\usepackage{mathtools}
\usepackage[output-decimal-marker={.}]{siunitx}
\usepackage{array}
\usepackage{tabularx}
\usepackage{multirow}
\usepackage{multicol}
\usepackage{setspace}
\usepackage{enumitem}
\usepackage{dblfloatfix}
\usepackage{wrapfig}
\usepackage{url}
\usepackage{hyperref}
\usepackage{cleveref}
\usepackage{eurosym}
\usepackage{tablefootnote}
\usepackage{comment}
\usepackage{ifthen}

\expandafter\def\expandafter\UrlBreaks\expandafter{\UrlBreaks
  \do\a\do\b\do\c\do\d\do\e\do\f\do\g\do\h\do\i\do\j%
  \do\k\do\l\do\m\do\n\do\o\do\p\do\q\do\r\do\s\do\t%
  \do\u\do\v\do\w\do\x\do\y\do\z\do\A\do\B\do\C\do\D%
  \do\E\do\F\do\G\do\H\do\I\do\J\do\K\do\L\do\M\do\N%
  \do\O\do\P\do\Q\do\R\do\S\do\T\do\U\do\V\do\W\do\X%
  \do\Y\do\Z}

\usetikzlibrary{positioning,fit,arrows.meta,backgrounds}
\usetikzlibrary{patterns,shapes,positioning}
\usetikzlibrary{decorations.markings}
\tikzset{
    module/.style={%
        draw, rounded corners,
        minimum width=#1,
        minimum height=8mm,
        font=\sffamily
        },
    module/.default=2.1cm,
    >=LaTeX
}
\usetikzlibrary{calc}

\newcommand\Tstrut{\rule{0pt}{2.6ex}}         
\newcommand\Bstrut{\rule[-0.9ex]{0pt}{0pt}}   

\pgfplotsset{
    compat = newest,
    layers/my layer set/.define layer set={b2,b1,main,f1,f2,f3}{},
    set layers=my layer set,
    /pgfplots/legend image code/.code={%
        \draw[mark repeat=2,mark phase=2,#1]
        plot coordinates {
        (0cm,0cm) 
        (0.15cm,0cm)
        (0.3cm,0cm)
        };%
    }
}

\allowdisplaybreaks

\begin{document}

\bstctlcite{IEEEexample:BSTcontrol}

\title{Towards Optimal Coordination between Regional Groups: HVDC Supplementary Power Control}%

\author{Andrea~Tosatto,~\IEEEmembership{Student~Member,~IEEE}, 
        Georgios~S.~Misyris,~\IEEEmembership{Student~Member,~IEEE}, 
        Adri{\`a}~Junyent-Ferr{\'e},~\IEEEmembership{Senior Member,~IEEE}, 
        Fei~Teng,~\IEEEmembership{Member,~IEEE},
        Spyros~Chatzivasileiadis,~\IEEEmembership{Senior Member,~IEEE}%
\thanks{A. Tosatto, G. S. Misyris and S. Chatzivasileiadis are with the Technical University of Denmark, Department of Electrical Engineering, Kgs. Lyngby, Denmark (emails: \{antosat,gmisy,spchatz\}@elektro.dtu.dk).}%
\thanks{A. Junyent-Ferr{\'e} and F. Teng are with the Department of Electrical and Electronic Engineering, Imperial College London, London SW7 2AZ, U.K. (email: \{adria.junyent-ferre,f.teng\}@imperial.ac.uk).}%
\thanks{This work is supported by Innovation Fund Denmark through the multiDC project (grant no. \mbox{6154-00020B}).}}

\maketitle

\vspace{-1.5em}
\begin{abstract}
With Europe dedicated to limiting climate change and greenhouse gas emissions, large shares of Renewable Energy Sources (RES) are being integrated in the national grids, phasing out conventional generation. The new challenges arising from the energy transition will require a better coordination between neighboring system operators to maintain system security. To this end, this paper studies the benefit of exchanging primary frequency reserves between asynchronous areas using the Supplementary Power Control (SPC) functionality of High-Voltage Direct-Current (HVDC) lines. First, we focus on the derivation of frequency metrics for asynchronous AC systems coupled by HVDC interconnectors. We compare two different control schemes for HVDC converters, which allow for unilateral or bilateral exchanges of reserves between neighboring systems. Second, we formulate frequency constraints and include them in a unit commitment problem to ensure the \mbox{N-1} security criterion. A data-driven approach is proposed to better represent the frequency nadir constraint by means of cutting hyperplanes. Our results suggest that the exchange of primary reserves through HVDC can reduce up to 10\% the cost of reserve procurement while maintaining the system N-1 secure.


\end{abstract}


\section*{Nomenclature}
\setlist[description]{font=\normalfont}
Below the list of the most important symbols in alphabetical order and grouped among indices, parameters, continuous variables and binary variables.
\vskip 0.5em
\noindent
\textsc{Indices and Sets:}
\vskip 0.2em
\begin{description}[leftmargin=3em,style=nextline, nosep]
\item[$a,\,b$] Asynchronous area index.
\item[$\mathcal{A}$] Set of asynchronous areas.
\item[$i$] Generator index.
\item[$\mathcal{G}$] Set of synchronous generators.
\item[$j$] Load index.
\item[$\mathcal{D}$] Set of loads.
\item[$k$] Converter index.
\item[$\mathcal{L}^{\textsc{ac}}$] Set of AC lines.
\item[$\mathcal{L}^{\textsc{dc}}$] Set of DC lines.
\item[$n,\,m$] Electrical node index.
\item[$\mathcal{N}$] Set of electrical nodes.
\item[$r$] RES producer index.
\item[$\mathcal{R}$] Set of RES producers.
\end{description}
\vskip 0.5em
\noindent
\textsc{Parameters:}
\vskip 0.2em
\begin{description}[leftmargin=3em,style=nextline, nosep]
\item[$\Delta f_a$] Instantaneous frequency deviation in area $a$.
\item[$\Delta f^{\rm ss}_a$] Steady-state frequency deviation in area $a$.
\item[$\Delta P_a$] Power deviation in area $a$.
\item[$B_{n,m}$] Susceptance of the line between bus $n$ and $m$.
\item[$C_i$] Linear production cost of generator $i$.
\item[$C^{su}_i$] Start-up cost of generator $i$.
\item[$C^{sd}_i$] Shut-down cost of generator $i$.
\item[$C^{r}_i$] Reservation cost of generator $i$.
\item[$\widehat{D}_a$] Generator damping in area $a$.
\item[$F^g_i$] Total fraction of power generated by the turbine of generator $i$.
\item[$\Dot{f}_a$] Rate of change of frequency in area $a$.
\item[$H_i$] Inertia constant of generator $i$.
\item[$I_{n,k}$] Incidence of HVDC line $k$ on bus $n$.
\item[$K^c_k$] Electric power gain factor of converter $k$.
\item[$K^g_i$] Mechanical power gain factor of generator $i$.
\item[$L_{j,t}$] Consumption of load $j$ at time $t$.
\item[$\overline{P}^{\textsc{ac}}_{n,m}$] Thermal limit of the AC line between bus $n$ and $m$.
\item[$\overline{P}^{\textsc{dc}}_{k}$] Thermal limit of the DC line $k$.
\item[$\overline{P}_i$] Maximum output of generator $i$.
\item[$\underline{P}_i$] Minimum output of generator $i$.
\item[$P^{\rm uw}_i$] Upward ramping limit of generator $i$.
\item[$P^{\rm dw}_i$] Downward ramping limit of generator $i$.
\item[$R^c_k$] Frequency drop gain of converter $k$.
\item[$R^g_i$] Frequency drop gain of generator $i$.
\item[$T^{\rm on}_i$] Minimum online duration of generator $i$.
\item[$T^{\rm off}_i$] Minimum offline duration of generator $i$.
\item[$T^c_k$] Time constant of converter $k$.
\item[$T^g_i$] Time constant of generator $i$.
\item[$V^{ll}_j$] Value of lost load for load $j$.
\item[$V^{cw}_r$] Value of curtailed wind for RES producer $r$.
\item[$W_{r,t}$] Output of RES producer $r$ at time $t$.
\end{description}
\vskip 0.5em
\noindent
\textsc{Continuous Variables:}\\
\noindent
All variables are defined for the time instance $t$, which is omitted in this list but always included throughout the paper. 
\vskip 0.2em
\begin{description}[leftmargin=3em,style=nextline, nosep]
\item[$\theta_n$] Voltage angle at bus $n$.
\item[$d^s_j$] Load shedding of load $j$.
\item[$\widetilde{F}^g_a$] Fraction of total power generated by the turbines in area $a$.
\item[$\widehat{F}_a$] Equivalent fraction of total power generated by the turbines in area $a$ with the inclusion of converters.
\item[$g_i$] Output of generator $i$.
\item[$g^s_j$] Frequency response of generator $i$.
\item[$\widehat{M}_a$] System inertia in area $a$.
\item[$p^{\textsc{dc}}_k$] power flow on HVDC line $k$.
\item[$p^s_k$] Frequency response of converter $k$.
\item[$\widetilde{R}^g_a$] Sum of generator droop gains in area $a$.
\item[$\widetilde{R}^c_{a,b}$] Sum of converter droop gains between area $a$ and $b$.
\item[$\widehat{R}_a$] Sum of droop gains in area $a$.
\item[$w^c_r$] Wind curtailment of RES producer $r$.
\end{description}
\vskip 0.5em
\noindent
\textsc{Binary Variables:}\\
\noindent
All variables are defined for the time instance $t$, which is omitted in this list but always included throughout the paper. 
\vskip 0.2em
\begin{description}[leftmargin=3em,style=nextline, nosep]
\item[$u_i$] Online status of generator $i$.
\item[$u^{\rm lcc}_k$] Flow direction on LCC-HVDC line $k$.
\item[$v^c_{a,b}$] Frequency support of area $b$ to area $a$.
\item[$v^g_i$] Participation to frequency regulation of generator $i$.
\item[$y_i$] Start-up of generator $i$.
\item[$z_i$] Shut-down of generator $i$.
\end{description}

\vspace{-0.5em}
\section{Introduction}\label{sec:1}
\IEEEPARstart{A}{fter} the Paris Agreement, the European Union has taken a leading role in the energy transition, with ambitious targets for energy efficiency (32.5\%) and RES penetration (32\%). 
With the high intermittence of RES and their location far from load centers, electricity flows are expected to increase and become more variable, requiring more efficient network development. 
In this regard, ENTSO-e has identified the need of more than 90 GW of new installed transmission capacity by 2030. Among the different projects, more than 60 GW are new High-Voltage Direct-Current (HVDC) lines \cite{intro_3}.

Additionally, large integration of inverter-based generation will result in decreasing kinetic energy (or inertia) in the system. With lower kinetic energy, Transmission System Operators (TSOs) could face difficulties in operating the system in a stable and reliable way \cite{intro_4}. These new challenges will require better coordination between regional entities, spanning from new enhanced operational processes to the prevention and management of common threats \cite{intro_2}.

Decrease of system inertia is classified as one of the biggest future challenges by system operators around the world. The problem is particularly pronounced in small systems, e.g. Ireland or Australia \cite{intro_5, intro_8}, but it is progressively growing to comprise also larger interconnected systems such as the Nordic countries (Denmark, Sweden, Norway and Finland) or the UK\cite{intro_9, intro_6}. As more and more system operators are facing these challenges, many technical and regulatory solutions have been proposed in the literature \cite{intro_10, intro_11, intro_12}, ranging from new balancing products and control methods to market measures. However, most of these solutions require a substantial change of paradigm, e.g. new grid codes and/or market rules, or expensive remedial actions, e.g. down-regulation of critical units, RES curtailment and load shedding. 
Recent studies have shown how the power set-points of HVDC converters can be adjusted to support the system in the event of power disturbances. Given the large number of HVDC lines connecting asynchronous areas (also called regional groups) in Europe (see \figurename~\ref{fig:1_map}), this method, referred to as Emergency Power Control (EPC) of HVDC links, is considered by ENTSO-e as one of the most promising among all the possible options \cite{intro_10}. In the literature, control strategies for enabling HVDC to participate in frequency support have been widely investigated \cite{martin1991modulation,du2007new,du2008comparison,miao2010wind,spallarossa2013influence,Huang2017,de2016stabilising,ambia2020adaptive,yu2014review,Mallada2016,jiang2017performance,bucurenciu2015frequency,junyent2014blending,zhu2012inertia}, and evidence of the economic benefits coming from the activation of the EPC functionality during low inertia events are presented in \cite{intro_14}. However, this functionality has never been analyzed in an optimization framework for the scheduling of frequency services. As result, the economic benefits of sharing primary frequency reserves between asynchronous areas during non-critical operation have, so far, never been determined.

\definecolor{RGCE}{rgb}{0.3412,0.4078,0.6471}%
\definecolor{RGN}{rgb}{0.5216,0.4824,0.6392}%
\definecolor{RGB}{rgb}{0.6627,0.4000,0.6000}%
\definecolor{RGUK}{rgb}{0.4706,0.5843,0.6588}%
\definecolor{RGI}{rgb}{0.4706,0.3922,0.3882}%

\begin{figure}[!t]
    \centering
    \resizebox{0.46\textwidth}{!}{%
    \begin{tikzpicture}
        
        \node[inner sep=0pt, anchor = south west] (network) at (0,0) {\includegraphics[trim = 0.2cm 0.65cm 0.65cm 0.65cm,clip,width=0.5\textwidth]{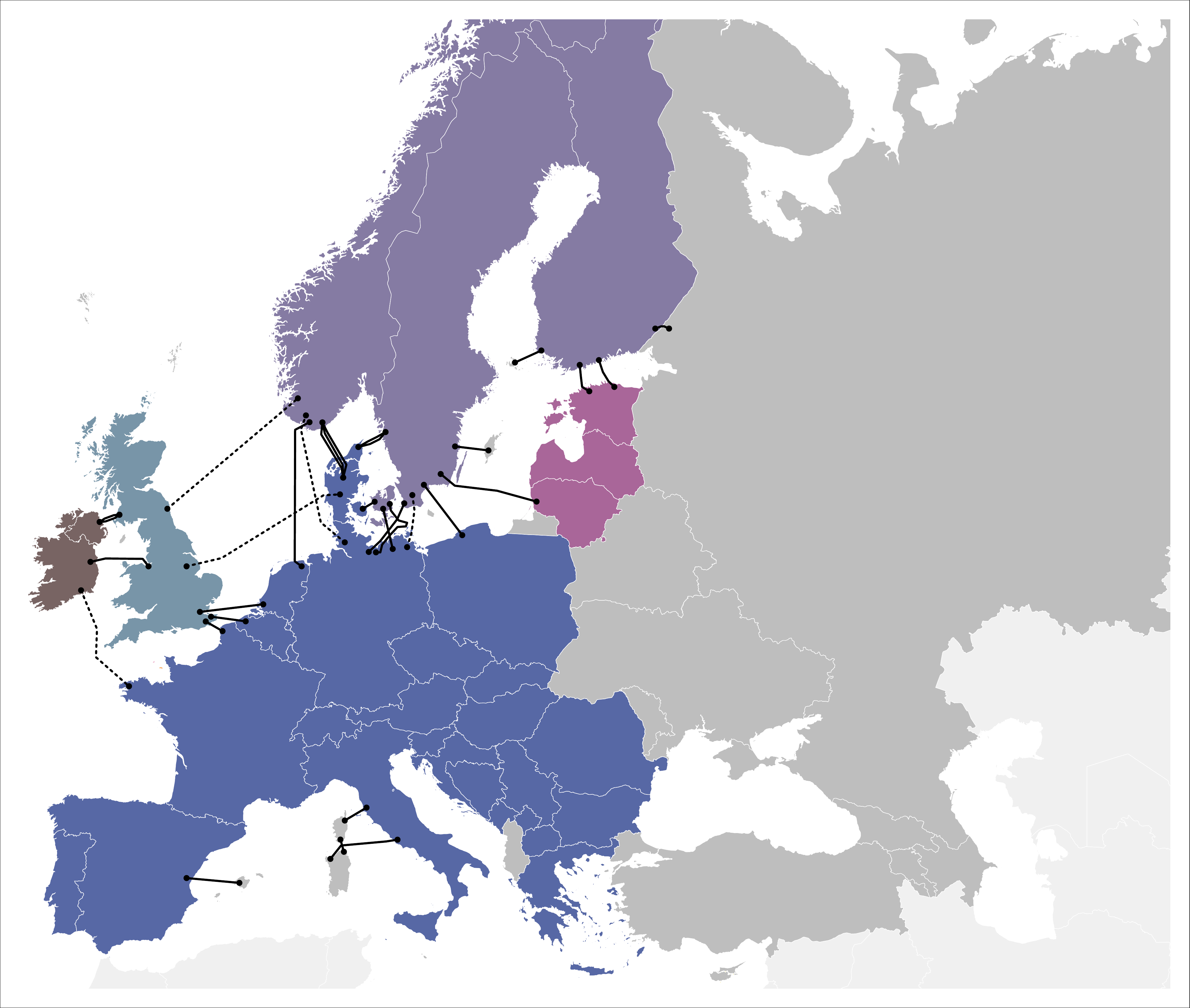}};
        \fill[fill=none, draw=black] (0.05,0) rectangle (9.08,7.54);
        \fill[fill=white, draw=black] (5.56,4) rectangle (8.86,7.32);
        \node[anchor=north west] at (5.56,7.3) {\raggedright\scriptsize \textbf{HVDC interconnectors:}};
        \fill (5.78,6.75) circle (0.03cm); 
        \fill (5.98,6.75) circle (0.03cm); 
        \draw[line width = 0.6,line cap=round] (5.78,6.755) -- (5.83,6.75); 
        \draw[line width = 0.6,line cap=round] (5.865,6.75) -- (5.895,6.75);
        \draw[line width = 0.6,line cap=round] (5.93,6.75) -- (5.98,6.75);
        \node[anchor=north west] at (6.06,6.95) {\raggedright\scriptsize Under construction}; 
        \fill (5.78,6.45) circle (0.03cm);
        \fill (5.98,6.45) circle (0.03cm);
        \draw[line width = 0.6] (5.78,6.45) -- (5.98,6.45);
        \node[anchor=north west] at (6.06,6.65) {\raggedright\scriptsize Operative};
        \node[anchor=north west] at (5.56,6.25) {\raggedright\scriptsize \textbf{Regional Groups (RG):}};
        \node[anchor=north west] at (6.06,5.9) {\raggedright\scriptsize RG Continental Europe};
        \draw[fill=RGCE,draw=none,line cap=round] (5.77,5.58) rectangle (5.99,5.78);
        \node[anchor=north west] at (6.06,5.55) {\raggedright\scriptsize RG Nordic};
        \draw[fill=RGN,draw=none,line cap=round] (5.77,5.23) rectangle (5.99,5.43);
        \node[anchor=north west] at (6.06,5.2) {\raggedright\scriptsize RG Baltic};
        \draw[fill=RGB,draw=none,line cap=round] (5.77,4.88) rectangle (5.99,5.08);
        \node[anchor=north west] at (6.06,4.85) {\raggedright\scriptsize RG UK};
        \draw[fill=RGUK,draw=none,line cap=round] (5.77,4.53) rectangle (5.99,4.73);
        \node[anchor=north west] at (6.06,4.5) {\raggedright\scriptsize RG Ireland};
        \draw[fill=RGI,draw=none,line cap=round] (5.77,4.18) rectangle (5.99,4.38);

    \end{tikzpicture}
    }\vspace{-0.5em}
    \caption{Regional groups in Europe and HVDC connections.}
    \label{fig:1_map}
    \vspace{-1.3em}
\end{figure}
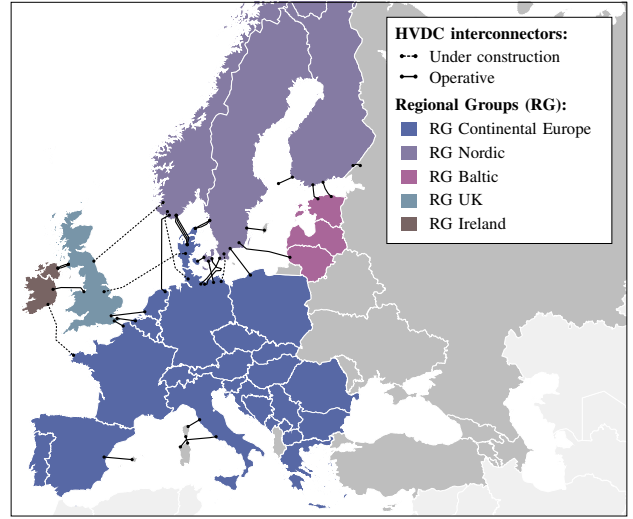

The aim of exchanging primary reserves via HVDC during normal operation (referred to as HVDC Supplementary Power Control - SPC) is to improve the frequency stability of the system while reducing the costs of reserve procurement. In order to have an optimization model responsive to the frequency behaviour of the system, frequency metrics, such as Rate of Change of Frequency (RoCoF), Instantaneous Frequency Deviation (IFD) and Steady-State Frequency Deviation (SSFD), must be included in the problem formulation. In this regard, two main challenges can be identified: (i) the analytical derivation of frequency metrics and (ii) the tractability of frequency constraints. Several works have already studied how to include frequency constraints in a Unit Commitment Problem (UCP) for systems with high RES penetration \cite{FMOD_1,FMOD_2,Badesa_2020,Chu2020}. In \cite{FMOD_1}, the authors derive analytical frequency metrics for RoCoF, IFD and SSFD and include them in a UCP by means of piecewise linear approximations. The authors in \cite{FMOD_2} and \cite{Badesa_2020}, extended the work of \cite{FMOD_1} by including converter control schemes and fast frequency response services, such as frequency droop and virtual synchronous machine controls provided by inverter-based generation units. The authors in \cite{Chu2020}, instead, present a UCP formulation which accounts for frequency support from variable speed wind turbines. Compared to \cite{FMOD_1}, \cite{FMOD_2} presents also a less computational intense method to limit the IFD, introducing safe bounds for the single variables. This method increases the tractability of the IFD constraint but disproportionately reduces the feasible space, calling for more efficient methods. Moreover, in all the aforementioned works the authors considered only one synchronous system, overlooking the option of exchanging primary frequency reserves among neighboring asynchronous systems via HVDC. 

To this end, this paper aims at deriving frequency metrics for asynchronous AC systems whose frequency dynamics are coupled by HVDC links, and to include the corresponding frequency constraints in an optimization framework to assess what the benefit of the HVDC SPC implementation is. To improve the accuracy and the computational efficiency of the proposed formulation, the IFD constraint is implicitly included in the optimization problem by limiting the feasible space with cutting hyperplanes. In detail, the contributions of this paper are:
\begin{itemize}
    \item The derivation of analytical expressions for the frequency metrics of asynchronous AC systems exchanging primary frequency services through HVDC links. 
    \item An improved unit commitment formulation with frequency constraints, which considers the dynamic response of generators and converters based on their droop coefficients;
    \item A data-driven approach to better represent the frequency nadir constraint by means of cutting hyperplanes;
    \item The inclusion of the SPC functionality of HVDC lines in an optimization framework for the exchange of primary reserves between neighboring asynchronous systems.
\end{itemize}

The remainder of this paper is structured as follows. \mbox{Section \ref{sec:2}} outlines two different control schemes for the implementation of HVDC SPC, and \mbox{Section \ref{sec:3}} describes the derivation of frequency metrics for asynchronous AC systems interconnected by HVDC links. \mbox{Section \ref{sec:4}} presents the optimization framework for the exchange of primary frequency services between asynchronous areas. A dedicated test case and simulation results are then described in \mbox{Section \ref{sec:5}}. Finally, \mbox{Section \ref{sec:6}} gathers conclusions and perspectives regarding further works.
\section{HVDC Supplementary Power Control Schemes}\label{sec:2}
Frequency controllers have been developed with the aim of balancing the active power in the system in the event of load disconnection or generator tripping. When implemented on HVDC converters, the outer control loop is modified to adjust the active power set-points based on the frequency deviation in the system \cite{Frandley2017}. A widely used approach for HVDC frequency control is the active power-frequency droop control method \cite{martin1991modulation,du2007new,du2008comparison,miao2010wind,spallarossa2013influence,Huang2017,de2016stabilising,ambia2020adaptive}. With this control scheme, the active power flow changes proportionally to the frequency deviation of the AC system, limiting the frequency nadir in the event of a power disturbance. An alternative approach is to use RoCoF measurements to react to a power disturbance. This control scheme is referred to as synthetic/virtual inertia and has been widely studied in \cite{yu2014review,Mallada2016,jiang2017performance}. The main advantage of this scheme is that it reduces the RoCoF and allows for inertia sharing between neighboring AC systems. However, as demonstrated in \cite{Mallada2016} the virtual inertia solution suffers from unbounded noise amplification when measurement noise is considered, which indicates that virtual inertia could potentially further degrade the grid performance once broadly deployed. Finally, a third alternative for inertia emulation control is the extraction of the stored energy in the DC link. This approach has been investigated in \cite{bucurenciu2015frequency,junyent2014blending,zhu2012inertia}; it uses input the frequency of the AC system to adjust the DC voltage reference. However, the limited energy stored in the HVDC link reduces the effectiveness of this control strategy. 

Therefore, in this paper, we consider HVDC frequency support based on droop frequency control, following the practice of most TSOs \cite{intro_10}. Moreover, we consider two control schemes, namely unilateral and bilateral, which determine whether HVDC converters react to the frequency deviation in only one area, or in both \cite{Huang2017}. In the following, these two schemes are presented. 

\vspace{-0.6em}
\subsection{Unilateral SPC Scheme}
In the unilateral control scheme (UCS), the frequency is measured only in one of the two interconnected AC systems. In the event of a disturbance in the monitored area, the active power exchange between the two areas is increased (or decreased) to balance the power mismatch. This means that part of the disturbance is propagated in the supporting area and some frequency reserves are activated to preserve the frequency in this area, which is equivalent to saying that some frequency reserves in the supporting area are exchanged with the area under contingency. A schematic representation of this control scheme is provided in \figurename~\ref{fig:2_ctrl}.

The modification of the active power set point of the HVDC converter following a frequency deviation can be expressed as
\begin{equation}\label{eq:unilateral}
\small 
\Delta P^{\rm ref}(s) = \frac{K^{c}}{R^{{c}}(1 + s T^{c})} \Delta f_a,
\end{equation}
with $P^{\rm ref}$ the active power reference, $K^{c}$, $R^{c}$ and $T^c$ respectively the electric power gain factor, the frequency droop value and the time constant of the active power controller of the converters, $s$ the Laplace operator and $\Delta f_a$ the frequency deviation in the monitored area (area $a$). 

The advantage of this scheme is that it reduces the IFD which follows a contingency in the supported area. However, the resulting frequency deviation in the supporting area is not monitored by the HVDC converter and only depends on the stiffness of the supporting AC system. In case a weak AC system provides frequency support to another asynchronous area, this could result in unacceptable large frequency excursions which in turn might force system operators in the supporting area to procure more frequency reserves.

\vspace{-0.6em}
\subsection{Bilateral SPC Scheme}

The bilateral control scheme (BCS) follows the same principles of the unilateral scheme, with the only difference that the frequency is measured in both the interconnected areas. The converter, thus, reacts to the frequency difference between the two areas and adjusts the active power flow accordingly. The active power set point of the HVDC converter is modified as follows:
\begin{equation}\label{eq:bilateral}
\small 
\Delta P^{\rm ref}(s) = \frac{K^{c}}{R^{{c}}(1+sT^{c})}\left(\Delta f_a - \Delta f_b \right)
\end{equation}
with $f_b$ the frequency of the supporting area (area $b$). Being responsive to the frequency deviations in both areas, the HVDC link can be used for supporting both areas. A schematic representation of this control scheme is also provided in \figurename~\ref{fig:2_ctrl}, with the differences between the two schemes marked in red.

Compared to the unilateral scheme, the main advantage of this scheme is that the exchange of frequency reserves is not limited to one direction. However, a change in the active power flow will cause a frequency deviation also in the second area, and thus the improvement of the IFD in the area under contingency is smaller.

\section{Frequency Dynamic Model}\label{sec:3}
In order to account for frequency stability issues in power systems and optimally dispatch synchronous generators, additional constraints must be included in the unit commitment problem to limit RoCoF, IFD and SSFD. Thus, in this section, we derive a simplified frequency response model of AC systems with thermal synchronous generators and HVDC interconnectors, as shown in \figurename~\ref{fig:2_ctrl}, which will be used for obtaining the frequency metrics. For the analytic formulation of the frequency metrics, we consider that a contingency cannot occur simultaneously in the interconnected AC systems. For the notation, generators are referred to with the index $i$, converters with the index $k$. When a parameter or variable refers to a specific area, e.g. the total inertia constant, the subscript refers to that area ($a$, $b$ and so on).

The dynamics that need to be taken into account for an accurate extraction of the center of inertia frequency of an AC system are: (i) the generator dynamics, (ii) governor droop and turbine dynamics and (iii) HVDC converter dynamics. The generator dynamics can be described by the swing equation (see \figurename~\ref{fig:2_ctrl}), where $\widehat{M}_a$ and $\widehat{D}_a$ are weighted system parameters representing the total inertia and damping constants of the generators in area $a$. A low-order model, proposed in \cite{Anderson1990} and evaluated in \cite{Shi2018}, is used to model the governor droops and turbine dynamics, with $R^{g}_i$ and $K^{g}_i$ respectively the frequency droop and mechanical power gain factor, $F^{g}_i$ the fraction of total power generated by the turbines of the synchronous generators and $T^{g}_i$ the turbine time constant. For the HVDC converters, the frequency droop controllers described by \eqref{eq:unilateral} and \eqref{eq:bilateral} are used to provide support during the containment phase, while frequency restoration is assumed to rely only on local reserves.

\definecolor{Blue1_CTRL}{rgb}{0.2863,0.3294,0.4471}%
\definecolor{Blue2_CTRL}{rgb}{0.8706,0.9216,0.9686}%
\definecolor{Blue3_CTRL}{rgb}{0.7412,0.8431,0.9333}%
\definecolor{Beige1_CTRL}{rgb}{0.5451,0.5529,0.4784}%
\definecolor{Beige2_CTRL}{rgb}{0.9098,0.9098,0.8941}%
\definecolor{Beige3_CTRL}{rgb}{0.8196,0.8196,0.7922}%

\usetikzlibrary {arrows.meta}

\begin{figure}[!t]
    \centering
    \resizebox{0.46\textwidth}{!}{%
    \begin{tikzpicture}
        
        \draw[fill=Blue2_CTRL, draw=Blue1_CTRL, rounded corners=13pt, thick] (0,5.4) rectangle (9,14);
        \draw[fill=Beige2_CTRL, draw=Beige1_CTRL, rounded corners=13pt, thick] (0,0.15) rectangle (9,5);
        \node[anchor=west] at (0.2,13.4) {\Large \textcolor{Blue1_CTRL}{\textbf{Area $\boldsymbol{a}$}}};
        \draw[fill=none, draw=Blue1_CTRL, rounded corners=13pt, thick, dashed] (0.3,9.3) rectangle (8,12.25);
        \node[anchor=west] at (0.5,11.125) {\textcolor{Blue1_CTRL}{\textsc{Turbine} \small \&}};
        \node[anchor=west] at (0.5,10.725) {\textcolor{Blue1_CTRL}{\textsc{Governor}}};
        \node[anchor=west] at (0.5,10.325) {\textcolor{Blue1_CTRL}{\textsc{Control}}};
        \draw[fill=none, draw=Blue1_CTRL, rounded corners=13pt, thick, dashed] (0.3,5.55) rectangle (8,8.5);
        \node[anchor=west] at (0.5,7.425) {\textcolor{Blue1_CTRL}{\textsc{HVDC}}};
        \node[anchor=west] at (0.5,7.025) {\textcolor{Blue1_CTRL}{\textsc{Converter}}};
        \node[anchor=west] at (0.5,6.625) {\textcolor{Blue1_CTRL}{\textsc{Control}}};
        
        \node[anchor=north west] at (2.8,13.5) {\small $\Delta P_a^{e}$};
        \node[anchor=north west] at (3.95,13.05) {\small $-$};
        \node[anchor=north west] at (8,13.5) {\small $\Delta f_a$};
        \node[anchor=north west] at (4.55,13.9) {\small\textcolor{Blue1_CTRL}{System dynamics}};
        \draw[fill=Blue3_CTRL, draw=black] (4.6,12.6) rectangle (7,13.4) node[pos=.5]{$\frac{1}{s\widehat{M}_a+\widehat{D}_a}$};
        \node[anchor=north west] at (4.8,12.3) {\small\textcolor{Blue1_CTRL}{1\textsuperscript{st} Generator}};
        \draw[fill=Blue3_CTRL, draw=black] (4.4,11) rectangle (7.2,11.8) node[pos=.5]{$\frac{K^g_{1}(1+sF^g_{1}T^g_{1})}{R^g_{1}(1+sT^g_{1})}$};
        \draw[fill=black, draw=black] (5.8,10.87) circle (0.01);
        \draw[fill=black, draw=black] (5.8,10.75) circle (0.01);
        \draw[fill=black, draw=black] (5.8,10.63) circle (0.01);
        \node[anchor=north west] at (4.7,10.7) {\small\textcolor{Blue1_CTRL}{M\textsuperscript{th} Generator}};
        \draw[fill=Blue3_CTRL, draw=black] (4.4,9.4) rectangle (7.2,10.2) node[pos=.5]{$\frac{K^g_{m}(1+sF^g_{m}T^g_{m})}{R^g_{m}(1+sT^g_{m})}$};
        \node[anchor=north west] at (4.35,8.55) {\small\textcolor{Blue1_CTRL}{1\textsuperscript{st} HVDC Converter}};
        \draw[fill=Blue3_CTRL, draw=black] (4.6,7.25) rectangle (7,8.05) node[pos=.5]{$\frac{K^c_{1}}{R^c_{1}(1+sT^c_{1})}$};
        \draw[fill=black, draw=black] (5.8,7.14) circle (0.01);
        \draw[fill=black, draw=black] (5.8,7.02) circle (0.01);
        \draw[fill=black, draw=black] (5.8,6.9) circle (0.01);
        \node[anchor=north west] at (4.3,6.95) {\small\textcolor{Blue1_CTRL}{N\textsuperscript{th} HVDC Converter}};
        \draw[fill=Blue3_CTRL, draw=black] (4.6,5.65) rectangle (7,6.45) node[pos=.5]{$\frac{K^c_{n}}{R^c_{n}(1+sT^c_{n})}$};
        
        \draw[->,-{Classical TikZ Rightarrow[length=3pt]}] (3,13)--(3.77,13);
        \draw[->,-{Classical TikZ Rightarrow[length=3pt]}] (4.03,13)--(4.6,13);
        \draw[->,-{Classical TikZ Rightarrow[length=3pt]}] (4.4,11.4)--(4.03,11.4);
        \draw[->,-{Classical TikZ Rightarrow[length=3pt]}] (4.4,9.8)--(4.03,9.8);
        \draw[->,-{Classical TikZ Rightarrow[length=3pt]}] (4.6,7.65)--(4.03,7.65);
        \draw[-] (4.6,6.05)--(3.9,6.05);
        \draw[fill=white, draw=black] (3.9,13) circle (0.13);
        \draw[fill=white, draw=black] (3.9,11.4) circle (0.13);
        \draw[fill=white, draw=black] (3.9,9.8) circle (0.13);
        \draw[fill=white, draw=black] (3.9,7.65) circle (0.13);
        \draw[->,-{Classical TikZ Rightarrow[length=3pt]}] (3.9,11.53)--(3.9,12.87);
        \draw[->,-{Classical TikZ Rightarrow[length=3pt]}] (3.9,9.93)--(3.9,11.27);
        \draw[->,-{Classical TikZ Rightarrow[length=3pt]}] (3.9,7.78)--(3.9,9.67);
        \draw[->,-{Classical TikZ Rightarrow[length=3pt]}] (3.9,6.05)--(3.9,7.52);
        
        \draw[->,-{Classical TikZ Rightarrow[length=3pt]}] (7,13)--(8.8,13);
        \draw[-,-{Classical TikZ Rightarrow[length=3pt]}] (7.7,13)--(7.7,9.03);
        \draw[-] (7.7,8.77)--(7.7,6.05);
        \draw[fill=black, draw=black] (7.7,13) circle (0.03);
        \draw[->,-{Classical TikZ Rightarrow[length=3pt]}] (7.7,11.4)--(7.2,11.4);
        \draw[fill=black, draw=black] (7.7,11.4) circle (0.03);
        \draw[->,-{Classical TikZ Rightarrow[length=3pt]}] (7.7,9.8)--(7.2,9.8);
        \draw[fill=black, draw=black] (7.7,9.8) circle (0.03);
        \draw[->,-{Classical TikZ Rightarrow[length=3pt]}] (7.7,7.65)--(7,7.65);
        \draw[fill=black, draw=black] (7.7,7.65) circle (0.03);
        \draw[->,-{Classical TikZ Rightarrow[length=3pt]}] (7.7,6.05)--(7,6.05);
        
        \draw[fill=black, draw=black] (3.9,8.9) circle (0.03);
        \draw[-] (3.9,8.9)--(3.3,8.9);
        \draw[-] (3.3,8.9)--(3.3,4);
        \draw[fill=white, draw=black] (7.7,8.9) circle (0.13);
        
        \draw[fill=none, draw=Beige1_CTRL, rounded corners=13pt, thick, dashed] (0.3,0.3) rectangle (8,3.25);
        \node[anchor=west] at (0.5,2.175) {\textcolor{Beige1_CTRL}{\textsc{Turbine} \small \&}};
        \node[anchor=west] at (0.5,1.775) {\textcolor{Beige1_CTRL}{\textsc{Governor}}};
        \node[anchor=west] at (0.5,1.375) {\textcolor{Beige1_CTRL}{\textsc{Control}}};
        \node[anchor=west] at (0.2,4.4) {\Large \textcolor{Beige1_CTRL}{\textbf{Area $\boldsymbol{b}$}}};
        \node[anchor=north west] at (2.4,4) {\small $\Delta P_b^{\textsc{hvdc}}$};
        \node[anchor=north west] at (3.95,4.05) {\small $-$};
        \node[anchor=north west] at (8,4) {\small $\Delta f_b$};
        \node[anchor=north west] at (4.55,4.9) {\small\textcolor{Beige1_CTRL}{System dynamics}};
        \draw[fill=Beige3_CTRL, draw=black] (4.6,3.6) rectangle (7,4.4) node[pos=.5]{$\frac{1}{s\widehat{M}_b+\widehat{D}_b}$};
        \node[anchor=north west] at (4.8,3.3) {\small\textcolor{Beige1_CTRL}{1\textsuperscript{st} Generator}};
        \draw[fill=Beige3_CTRL, draw=black] (4.4,2) rectangle (7.2,2.8) node[pos=.5]{$\frac{K^g_{1}(1+sF^g_{1}T^g_{1})}{R^g_{1}(1+sT^g_{1})}$};
        \draw[fill=black, draw=black] (5.8,1.87) circle (0.01);
        \draw[fill=black, draw=black] (5.8,1.75) circle (0.01);
        \draw[fill=black, draw=black] (5.8,1.63) circle (0.01);
        \node[anchor=north west] at (4.8,1.7) {\small\textcolor{Beige1_CTRL}{X\textsuperscript{th} Generator}};
        \draw[fill=Beige3_CTRL, draw=black] (4.4,0.4) rectangle (7.2,1.2) node[pos=.5]{$\frac{K^g_{x}(1+sF^g_{x}T^g_{x})}{R^g_{x}(1+sT^g_{x})}$};
        
        \draw[->,-{Classical TikZ Rightarrow[length=3pt]}] (3.3,4)--(3.77,4);
        \draw[->,-{Classical TikZ Rightarrow[length=3pt]}] (4.03,4)--(4.6,4);
        \draw[fill=white, draw=black] (3.9,4) circle (0.13);
        \draw[fill=white, draw=black] (3.9,2.4) circle (0.13);
        \draw[->,-{Classical TikZ Rightarrow[length=3pt]}] (4.4,2.4)--(4.03,2.4);
        \draw[-] (4.4,0.8)--(3.9,0.8);
        \draw[->,-{Classical TikZ Rightarrow[length=3pt]}] (3.9,2.53)--(3.9,3.87);
        \draw[->,-{Classical TikZ Rightarrow[length=3pt]}] (3.9,0.8)--(3.9,2.27);
        
        \draw[->,-{Classical TikZ Rightarrow[length=3pt]}] (7,4)--(8.8,4);
        \draw[-] (7.7,4)--(7.7,0.8);
        \draw[fill=black, draw=black] (7.7,4) circle (0.03);
        \draw[->,-{Classical TikZ Rightarrow[length=3pt]}] (7.7,2.4)--(7.2,2.4);
        \draw[fill=black, draw=black] (7.7,2.4) circle (0.03);
        \draw[->,-{Classical TikZ Rightarrow[length=3pt]}] (7.7,0.8)--(7.2,0.8);
        
        \node[anchor=north west] at (7.75,8.9) {\small \textcolor{red}{$-$}};
        \draw[->,-{Classical TikZ Rightarrow[length=3pt]},draw=red] (8.3,8.9)--(7.83,8.9);
        \draw[-,draw=red] (8.3,4)--(8.3,8.9);
        \draw[fill=red, draw=red] (8.3,4) circle (0.03);

    \end{tikzpicture}
    }%
\caption{System dynamics model under unilateral and bilateral HVDC SPC schemes. The difference between the two schemes is highlighted in red. In the case of the unilateral scheme, the line in red is disregarded, which means that the active power exchanged with the grid is only dependent on the frequency of the AC grid under the contingency. In the case of the bilateral scheme, the line in red is taken into account, which means that the active power adjustment is dependent on both frequencies of the interconnected AC grids.}
\label{fig:2_ctrl}
\vspace{-1em}
\end{figure}
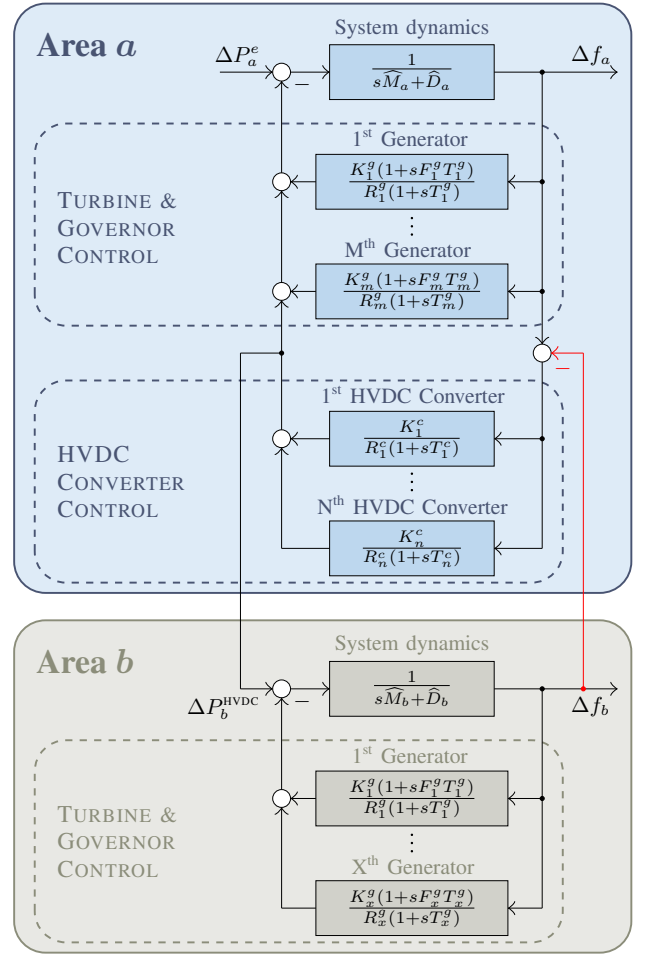

\subsection{Frequency Metrics - Unilateral Scheme}
The relation between an active power disturbance $\Delta P^e_a$ and the frequency deviation $\Delta f_a$ in the supported area (area $a$ from now on) is given by:
\begin{equation}\label{Eq:UniClosedLoop}
\footnotesize
\begin{split}
    G^{\textsc{u}}_a(s) = \frac{\Delta f_a(s)}{\Delta P^e_a(s)} = &\Bigg(\Big(s \widehat{M}_a + \widehat{D}_a\Big) + \sum_{i \in \mathcal{G}^S_a}\frac{K^g_{i}\left(1+sF^g_{i}T^g_{i}\right)}{R^g_{i}\left(1+ sT^g_{i}\right)} \\
    & + \sum_{k \in \mathcal{L}^S_a}\frac{K^c_{k}}{R^c_{k}\left(1+ sT^c_{k}\right)}\Bigg)^{-1},
\end{split}
\normalsize
\end{equation}
with $\mathcal{G}^S_a$ and $\mathcal{L}^S_a$ respectively the set of generators and HVDC interconnectors providing frequency support in area $a$. Similar to \cite{Markovic2019}, we consider that the turbine time constants are equal, i.e. $T^{g}_i=T_a$, for all the synchronous machines within the same synchronous area and that the converter time constants are neglected ($T^{g} \gg T^{c}$). With these assumptions, \eqref{Eq:UniClosedLoop} takes the following generalized form:
\begin{equation}\label{Eq:genUCSa}
\small
    G^{\textsc{u}}_a(s) = \frac{1}{\widehat{M}_a T_a} \frac{1+sT_a}{s^2 + 2\zeta_a \omega_a^{n} s +{\omega_a^n}^2},
\end{equation}
where the natural frequency $\omega_a^n$ and the damping ratio $\zeta_a$ are given by:
\begin{equation}
\footnotesize
    \omega_a^n = \sqrt{\frac{\widehat{D}_a + \widehat{R}_a}{\widehat{M}_aT_a}}, \quad\quad \zeta_a = \frac{\widehat{M}_a + T_a(\widehat{D}_a+\widehat{F}_a)}{2\sqrt{\widehat{M}_aT_a(\widehat{D}_a + \widehat{R}_a)}}.
\end{equation}
$\widehat{M}_a$, $\widehat{D}_a$, $\widehat{F}_a$ and $\widehat{R}_a$ are weighted model parameters, obtained by aggregating system parameters and parameters of generators and HVDC converters in area $a$. For their derivation, a detailed mathematical formulation is presented in \cite{Markovic2019}. Similar to \cite{intro_13}, the damping constant $\widehat{D}_{a}$ is considered a fixed parameter. By applying the inverse Laplace transformation and assuming a step-wise disturbance $\Delta P_a$, we can derive a time-domain expression of the frequency deviation. This enables to extract the RoCoF, the IFD and the SSFD in area $a$, respectively:

\begin{subequations}
\footnotesize\vspace{-1em}
\begin{align}
    \Dot{f}_a^{\rm max} &= -\frac{\Delta P_a}{\widehat{M}_a}, \label{eq:rocof_unil_a}\\
    \Delta f_a^{\rm max} &= -\frac{\Delta  P_a}{\widehat{D}_a+\widehat{R}_a}\left(1 + \sqrt{\frac{T_a(\widehat{R}_a - \widehat{F}_a)}{\widehat{M}_a}}e^{-\zeta_a\omega_a^n t^m_a}\right), \label{eq:ifd_unil_a}\\
    \Delta f_a^{\rm ss} &= -\frac{\Delta P_a}{\widehat{D}_a + \widetilde{R}^g_a}, \label{eq:ssfd_unil_a}
\end{align}
\end{subequations}
where $t_a^{m}$ is the time instant when the maximum frequency deviation occurs, i.e. $\Delta \Dot{f}_a(t_a^{m})=0$. 
The obtained expressions are similar to the ones derived in \cite{FMOD_2}, meaning that HVDC converters act as generators and provide the necessary power to contain the frequency in area $a$. After the frequency containment phase, instead, only generators provide frequency support, thus only the term $\widetilde{R}^g_a$ (which is the sum of the droop gains of generators in area $a$) appears in \eqref{eq:ssfd_unil_a}.


In case the supporting TSOs want to limit the frequency deviation in their areas, additional constraints could be included. In the supporting area (area $b$ from now on), the change in the converters set-point $\Delta P^\textsc{hvdc}_b$ corresponds to the frequency deviation $\Delta f_b(s)$. The transfer function of area $b$, which is the mapping of the Laplace transform of $\Delta f_a(s)$ to the Laplace transform of $\Delta f_b(s)$, can be derived in a similar way to \eqref{Eq:genUCSa}, resulting in:
\begin{equation}\label{Eq:genUCSb}
\small
\begin{split}
     G^{\textsc{u}}_b(s) = \frac{\Delta f_b(s)}{\Delta f_a(s)} = \frac{\widetilde{R}^{c}_{a,b}}{\widehat{M}_b T_b} \frac{1+sT_b}{s^2 + 2\zeta_b \omega_b^n s +{\omega_b^n}^2}, 
\end{split}
\end{equation}
Since this area is not receiving support from area $a$ through HVDC, $\omega_b^n$ and $\zeta_b^n$ are calculated by replacing $\widehat{R}_b$ and $\widehat{F}_b$ respectively with $\widetilde{R}^g_b$ and $\widetilde{F}^g_b$, meaning that only the generators in area $b$ (and not converters) provide frequency support. The term $\widetilde{R}^{c}_{a,b}$, instead, refers only to the contribution of the HVDC links connecting area $a$ and $b$. 
By applying the inverse Laplace transformation and assuming a step-wise disturbance $\Delta f_a(s) = \Delta f^{\rm max}_a/s$, we can derive a time-domain expression of the frequency deviation in area $b$. This enables to extract frequency metrics for the supporting area as well:

\begin{subequations}
\footnotesize\vspace{-1em}
\begin{align}
    \Dot{f}^{\rm max}_b &= -\frac{\Delta P_a \widetilde{R}^{c}_{a,b}}{\widehat{M}_a\widehat{M}_b}, \label{Eq:Uni_dfmax_j} \\
    \Delta f^{\rm max}_b &= \frac{\Delta f^{\rm max}_a\widetilde{R}^{c}_{a,b}}{\widehat{D}_b+\widehat{R}_b}\left(1 + \sqrt{\frac{T_b(\widehat{R}_b - \widehat{F}_b)}{\widehat{M}_b}}e^{-\zeta_b\omega_b^n t^m_b}\right),  \label{Eq:Uni_fmax_j} 
\end{align}
\end{subequations}
It should be noted that, by considering $\Delta f_a(s)$ as a step-wise disturbance, it is not possible to calculate the exact solution for $\Delta f_b(t)$ since its trajectory varies based on the generator, turbine governor and HVDC converters parameters. Although this results in the wrong calculation of the time to nadir, $t^m_b$, $\Delta f^{\rm max}_b$ is accurately represented. Moreover, the SSFD in the supporting area is equal to zero as there is no contribution of HVDC to frequency restoration in area $a$.  


\vspace{-0.5em}
\subsection{Frequency Metrics - Bilateral Scheme}
The main difference between the bilateral and the unilateral schemes is that, in the bilateral scheme, the active power set-points of the HVDC converters change proportionally to the difference between the frequencies of the interconnected AC systems. In the event of a power disturbance $\Delta P_a$, the corresponding frequency deviation is given by:
%

\vspace{-1em}
\begin{equation} \label{Eq:genBCSa}
    \footnotesize
    \Delta f_a(s) = G^{\textsc{u}}_a(s)\Bigg(\Delta P^e_a(s) + \sum_{k \in \mathcal{C}^S_a}\frac{K^c_k}{R^c_k\left(1+ sT^c_k\right)} \Delta f_b(s) \Bigg),
\end{equation}
showing that the frequency deviation in the area under contingency depends also on the frequency deviation in the supporting area. To derive expressions of the frequency metrics, first we need to analyze the frequency deviation in \mbox{area $b$}. The transfer function $G^{\textsc{b}}_b(s)$, which models the relation between $\Delta f_b(s)$ and $\Delta f_a(s)$, is similar to \eqref{Eq:genUCSb}, where $\omega_b^n$ and $\zeta_b$ can be calculated using $\widehat{R}_b$ and $\widehat{F}_b$, meaning that both generators and HVDC converters participate in frequency support in area $b$.
%

Plugging $G^{\textsc{b}}_b(s)$ into \eqref{Eq:genBCSa} introduces additional complexity to the derivation of the analytical expressions, since (even after the assumptions made in the previous sections) the transfer function $\Delta f_b(s)/\Delta P_a^e(s)$ becomes a fourth-order transfer function and, in turn, $\Delta f_a(s)/\Delta P_a^e(s)$ becomes a six-order transfer function. To overcome this issue, similarly to the previous section, we first consider a step-wise change of $\Delta f_a(s) = \Delta f_a^{\rm max}/s$, where $\Delta f_a^{\rm max}$ is the maximum allowable frequency deviation determined by the transmission system operator in area $a$. This assumption results in an overestimation of the absolute value of the frequency nadir in both areas. However, this guarantees that the RoCoF and the IFD stay within the operational limits. The frequency metrics for area $b$ are similar to the ones described by \eqref{Eq:Uni_dfmax_j} and \eqref{Eq:Uni_fmax_j}, where $\widehat{R}_b$ and $\widehat{F}_b$ consider both generators and HVDC converters participating in the frequency support of area $b$. 

Similar to the UCS case, we consider that there is no contribution of HVDC to the SSFD of the AC network that faces the contingency. To derive the frequency metrics for area $a$, we consider stepwise changes $\Delta P_a^e(s)$ and $\Delta f_a$ as $-\Delta P_a/s$ and $\Delta f_a^{max}/s$, respectively. This leads to similar expressions for the RoCoF and steady state frequency deviation as in \eqref{eq:rocof_unil_a} and \eqref{eq:ssfd_unil_a} and the following expression for the frequency nadir:
\begin{equation}
\footnotesize
    \Delta f^{\rm max}_a = \frac{-\Delta P_a - \Delta f_{b}^{\rm max}\widetilde{R}^{c}_{a,b}}{\widehat{D}_b+\widehat{R}_{a}}\left(1 + \sqrt{\frac{T_a(\widehat{R}_{a} - \widehat{F}_{a})}{\widehat{M}_a}}e^{-\zeta_a\omega_{a}^n t^m_a}\right) \label{eq:ifd_bil_a}.
\end{equation}
We remind again the reader that \eqref{eq:ifd_bil_a} overestimates the value of the frequency nadir. However, it still ensures that the frequency of the area $a$ will remain within the limits set by the transmission system operators.
\section{Optimization Framework}\label{sec:4}
The calculation of the weighted model parameters depends on which synchronous generators are online and able to provide frequency support. Thus, in order to include frequency constraints in an optimization problem, specific variables associated with the status of the units must be defined. In a Unit Commitment Problem (UCP), the status of generators is modelled by means of binary variables. This allows to include ramping limits, minimum and maximum generation limits, minimum online and offline duration and start-up and shut-down costs. The frequency metrics derived in the previous section can thus be directly included in a UCP without modifying its basic structure. In this section, we present the UCP formulation, augmented with frequency constraints and reserve sharing through HVDC.

\vspace{-0.5em}
\subsection{Objective Function}
The objective function contains the sum of all the costs associated with power generation, reserve procurement, load shedding and wind curtailment:
\begin{equation}\label{eq:obj}
\footnotesize
\begin{split}
    \sum_{t\in\mathcal{T}} & \sum_{i\in\mathcal{G}} \big(C_ig_{i,t}+C^{su}_i y_{i,t}+C^{sd}_i z_{i,t}+C^{r}_i g^s_{i,t}\big) \\
    & + \sum_{t\in\mathcal{T}} \sum_{j\in\mathcal{D}} \big(V^{ll}_j d^{s}_{j,t}\big) + \sum_{t\in\mathcal{T}} \sum_{r\in\mathcal{R}} \big(V^{cw}_r w^{c}_{r,t}\big)
\end{split}
\normalsize
\end{equation}
with $C_i$, $C^{su}_i$, $C^{sd}_i$ and $C^{r}_i$ respectively the production, start up, shut down and reservation costs of unit $i$, $V^{ll}_j$ the value of lost load of consumer $j$ and $V_r^{cw}$ the value of curtailed wind of producer $r$. The continuous variables $g_{i,t}$, $g^s_{i,t}$, $d^{s}_{j,t}$ and $w^{c}_{r,t}$ refer to the output and reserve of unit $i$, the amount of load shedding of consumer $j$ and the amount of wind curtailed for producer $r$. The binary variables $y_{i,t}$ and $z_{i,t}$ model the start-up and shut-down of unit $i$.

\vspace{-0.5em}
\subsection{Generator Constraints}
Generator limits are enforced by adding this set of constraints:

\begin{subequations}
\small\vspace{-1em}
\begin{alignat}{2}
    & u_{i,t} \underline{P}_i \leq g_{i,t} \leq u_{i,t} \overline{P}_i \quad\quad && \forall i, \forall t \label{eq:gen1}\\
    & g_{i,t}-g_{i,t-1} \leq P^{\rm uw}_i \quad\quad && \forall i, \forall t \\
    & g_{i,t-1}-g_{i,t} \leq P^{\rm dw}_i \quad\quad && \forall i, \forall t \\
    & y_{i,t} - \textstyle\sum_{\tau=t+1}^{\tau_i^{\rm on}} z_{i,\tau} \leq 1 \quad\quad && \forall i, \forall t \\
    & z_{i,t} - \textstyle\sum_{\tau=t+1}^{\tau_i^{\rm off}} y_{i,\tau} \leq 1 \quad\quad && \forall i, \forall t \\
    & y_{i,t} \geq u_{i,t}-u_{i,t-1} \quad\quad && \forall i, \forall t \label{eq:startup}\\
    & z_{i,t} \geq u_{i,t-1}-u_{i,t} \quad\quad && \forall i, \forall t \label{eq:shutdown}
\end{alignat}
\end{subequations}
with $\underline{P}_i$ and $\overline{P}_i$ the minimum and maximum generation output, $P^{\rm uw}_i$ and $P^{\rm dw}_i$ the upward and downward ramping limits, $\tau_i^{\rm on} = \text{min}\{t+T^{\rm on}_i-1,T\}$, $\tau_i^{\rm off} = \text{min}\{t+T^{\rm off}_i-1,T\}$, $T^{\rm on}$ and $T^{\rm off}$ the minimum online and offline duration and $T$ the last time instance considered in the problem. The binary variable $u_{i,t}$ defines the online or offline status of unit $i$ at the time instance $t$, and constraints \eqref{eq:startup}-\eqref{eq:shutdown} define the start-up and shut-down variables accordingly.

\vspace{-0.5em}
\subsection{Network and System Constraints}
Flows over AC and HVDC lines are constrained by the thermal capacity of conductors and, for the latter, power electronics devices. In addition, the change of polarity of a Line-Commutated Converter (LCC) requires de-energizing the converter station; therefore, contrary to Voltage-Source Converters (VSCs), it is not possible to rapidly change the direction of power flow \cite{LESAGELANDRY2020}. These constraints are defined as:

\begin{subequations}
\small\vspace{-1em}
\begin{alignat}{2}
    & B_{n,m}(\theta_{n,t}-\theta_{m,t}) \leq \overline{P}^{\textsc{ac}}_{n,m} \quad\quad && \forall (n,m), \forall t \label{eq:ac_lim}\\
    & -\overline{P}^{\textsc{dc}}_{k} \leq p^{\textsc{dc}}_{k,t} \leq \overline{P}^{\textsc{dc}}_{k} \quad\quad && \forall k\in\mathcal{L}^{\rm vsc}, \forall t \label{eq:dc_limVSC}\\
    & -(1-u^{\rm lcc}_k)\overline{P}^{\textsc{dc}}_{k} \leq p^{\textsc{dc}}_{k,t} \leq u^{\rm lcc}_k\overline{P}^{\textsc{dc}}_{k} \quad\quad && \forall k\in\mathcal{L}^{\rm lcc}, \forall t \label{eq:dc_limLCC}
\end{alignat}
\end{subequations}
with $\overline{P}^{\textsc{ac}}_{n,m}$ and $\overline{P}^{\textsc{dc}}_{k}$ the transmission capacity of AC and HVDC lines respectively, $B_{n,m}$ the susceptance of the line between bus $n$ and $m$, $\theta_{n,t}$ the voltage angle at bus $n$, $\mathcal{L}^{\rm vsc}$ and $\mathcal{L}^{\rm lcc}$ respectively the set of VSC- and LCC-based HVDC links (\mbox{$\mathcal{L}^{\rm vsc}\cup\mathcal{L}^{\rm lcc}=\mathcal{L}^{\textsc{dc}}$}), $p^{\textsc{dc}}_{k,t}$ the flow over the HVDC line $k$ and $u^{\rm lcc}_k$ the binary variable defining the flow direction on LCC-HVDC links. By using a binary variable in Eq. \eqref{eq:dc_limLCC}, the flow can be constrained to one direction for a predefined amount of time, e.g. 4 hours or 1 day.  

The nodal balance equation is expressed as:
\begin{equation}\label{eq:bal}
\footnotesize
    \begin{split}
        -\sum_{i\in\mathcal{G}_n} & g_{i,t} - \sum_{r\in\mathcal{R}_n} \big(W_{r,t}-w^{c}_{r,t}\big) + \sum_{j\in\mathcal{D}_n} \big(L_{j,t}-d^{s}_{j,t}\big) \\
        & + \sum_{m:(n,m)\in\mathcal{L}^{\textsc{ac}}} B_{n,m}(\theta_{n,t}-\theta_{m,t}) + \sum_{k\in\mathcal{L}^{\textsc{dc}}} I_{n,k}p^{\textsc{dc}}_{k,t} = 0 
    \end{split}
\normalsize
\end{equation}
with $\mathcal{G}_n$, $\mathcal{R}_n$ and $\mathcal{D}_n$ respectively the sets of generators, RES and loads connected to node $n$, $\mathcal{L}^{\textsc{ac}}$ and $\mathcal{L}^{\textsc{dc}}$ the sets of AC and HVDC lines respectively, $W_{r,t}$ and $L_{j,t}$ the RES production (producer $r$) and load consumption (consumer $j$) at bus $n$, and $I_{n,k}$ the coefficient of the HVDC incidence matrix corresponding to line $k$ and bus $n$.

\input{Plots/plot_2Dplane.tex}



\vspace{-0.5em}
\subsection{Weighted Model Parameters}
The participation of generators and HVDC converters to frequency support is decided by means of additional binary variables: $v^g_{i,t}$ defines the participation of generators while $v^c_{a,b,t}$ the areas that are supported (index $a$) or support (index $b$). Since only online generators can offer frequency support, the following constraint is enforced:
\begin{equation}\label{eq:v_gen}
    v^g_{i,t} \leq u_{i,t} \quad\quad \forall i, \forall t
\end{equation}
For the HVDC converters, depending on the SPC scheme applied, two different constraints are introduced. For the unilateral scheme:
\begin{equation}\label{eq:v_uni}
    v^c_{a,b,t} + v^c_{b,a,t} \leq 1 \quad\quad \forall a,\,\forall b\in\mathcal{A}	\setminus\{a\},\, \forall t 
\end{equation}
with $\mathcal{A}$ the set of asynchronous areas. Eq. \eqref{eq:v_uni} enforces that if area $a$ is supported by area $b$, area $a$ cannot support area $b$. 
For the bilateral scheme, instead, the following constraint is introduced:
\begin{equation}\label{eq:v_bi}
    v^c_{a,b,t} = v^c_{b,a,t} \quad\quad \forall a,\,\forall b\in\mathcal{A}	\setminus\{a\},\, \forall t 
\end{equation}
meaning that if area $a$ is supported by area $b$, also area $b$ is supported by area $a$.

With the new variables, the weighted model parameters are calculated as follows:
\begin{subequations}
\begin{alignat}{2}
    & \widetilde{R}^g_{a,t} = \textstyle\sum_{i\in\mathcal{G}_a} \tfrac{K^g_i}{R^g_i}\overline{P}_iv^g_{i,t} \quad\quad && \forall a,\,\forall t \label{eq:R_g}\\
    & \widetilde{R}^c_{a,t} = \textstyle\sum_{k\in\mathcal{L}^{\textsc{dc}}_a} \tfrac{K^c_k}{R^c_k}\overline{P}^{\textsc{dc}}_kv^c_{a,\sim,t} \quad\quad && \forall a,\,\forall t \label{eq:Rctot}\\
    & \widetilde{F}^g_{a,t} = \textstyle\sum_{i\in\mathcal{G}_a} \tfrac{K^g_iF^g_i}{R^g_i}\overline{P}_iv^g_{i,t} \quad\quad && \forall a,\,\forall t \\
    & \widehat{R}_{a,t} = \widetilde{R}^g_{a,t} + \widetilde{R}^c_{a,t} && \forall a,\,\forall t \\
    & \widehat{F}_{a,t} = \widetilde{F}^g_{a,t} + \widetilde{R}^c_{a,t} && \forall a,\,\forall t \\
    & \widehat{M}_{a,t} = \textstyle\sum_{i\in\mathcal{G}_a} 2H^g_i\overline{P}_iu^g_{i,t} && \forall a,\,\forall t \label{eq:M}
\end{alignat}
\end{subequations}
with $H^g_i$ the inertia constant of the synchronous unit $i$. Note that the subscript $\sim$ in \eqref{eq:Rctot} means whatever area is connected to area $a$ through the HVDC line $k$.

\vspace{-0.5em}
\subsection{Frequency Constraints}
The weighted model parameters are calculated to include frequency constraints in the optimization problem. This is done in order to limit the RoCoF, the IFD and the SSFD in the event of the worst contingency (N-1 security criterion), i.e. the loss of the biggest generating unit. For this, we consider a fixed power deviation $\Delta P^e_a = \text{max}\{\overline{P}_i\}$, with $i\in\mathcal{G}_a$. This assumption is justified by the fact that the biggest generating units are usually nuclear power plants (mostly) contracted in the long term market to always produce at full capacity. Moreover, we do not consider these events to happen simultaneously in more than one area.

From \eqref{eq:rocof_unil_a}, it is possible to observe a linear dependence of the RoCoF on the inertia level of the AC system. Similarly, \eqref{eq:ssfd_unil_a} shows a linear dependence between the total sum of droop coefficients and the SSFD. Thus, these two frequency metrics can be bounded by bounding the corresponding weighted model parameters:

\begin{subequations}
\small\vspace{-1em}
\begin{alignat}{2}
    & \widehat{M}_{a,t} \geq \frac{f^{\rm base}_a}{\Dot{f}^{\rm max}_a} \Delta P^e_a \quad\quad && \forall a,\,\forall t \label{eq:rocof}\\
    & \widetilde{R}^g_{a,t} \geq \frac{f^{\rm base}_a}{\Delta f^{\rm ss,max}_a} \Delta P^e_a - \widehat{D}_{a,t}  \quad\quad && \forall a,\,\forall t \label{eq:ssfd}
\end{alignat}
\end{subequations}
Being $\widehat{R}_{a,t}$ a linear combination of $\widetilde{R}^g_{a,t}$ and $\widetilde{R}^c_{a,t}$, by bounding $\widetilde{R}^g_{a,t}$ also $\widehat{R}_{a,t}$ is automatically bounded.

Regarding the IFD, it is not possible to directly include \eqref{eq:ifd_unil_a} and \eqref{eq:ifd_bil_a} in the optimization problem as they are non-linear and non-convex. The authors in \cite{FMOD_1} introduce a piecewise linearization of the frequency nadir which consists of a set of hyperplanes, activated by binary variables, that are included as equality constraints. The authors in \cite{FMOD_2}, instead, introduce bounds on the single variables to avoid excursions of the frequency beyond the limits. However, the first approach requires many hyperplanes to accurately represent the nadir function, making the problem computationally intense to solve, while the second significantly reduces the feasible space. Within reasonable bounds (determined by realistic values of the weighted model parameters), \eqref{eq:ifd_unil_a} is monotonically decreasing in all the considered variables, meaning that the higher is $\widehat{R}_{a,t}$, $\widehat{F}_{a,t}$ or $\widehat{M}_{a,t}$, the smaller is the IFD (see \figurename~\ref{fig:hyperplane}). Thus, the points on the boundary (i.e. close to the IFD limit) can be approximated with a hyperplane. This hyperplane, which is a linear combination of the weighted model parameters, divides the feasible space into two regions, one with the triplets ($\widehat{R}_{a,t}$,$\widehat{F}_{a,t}$,$\widehat{M}_{a,t}$) that violate the IFD requirements and one with the ones that do not. Thus in order to discard the triplets that violate the IFD constraints, the following inequality constraint is included in the optimization problem (for the UCS case):
\begin{equation}\label{eq:nadir_uni}
\small
    \widehat{F}_{a,t} \geq A^{\textsc{R}}_a \widehat{R}_{a,t} + A^{\textsc{M}}_a \widehat{M}_{a,t} + A^0_a \quad\quad \forall a,\,\forall t 
\end{equation}
with $A^{\textsc{R}}_a$, $A^{\textsc{M}}_a$ and $A^0_a$ the coefficients of the hyperplane. More information about the calculation of these coefficients is provided in Appendix \ref{appendix:A}. The left 3D plot of \figurename~\ref{fig:hyperplane} shows an example of hyperplane (depicted in orange) that divides the points that do not satisfy the IFD constraint (yellow and light green points below the plane) from the ones that satisfy it (dark green and blue points above the plane). Indeed, the colors assigned to each point are related to the value of IFD for the corresponding system parameters $\widehat{R}_{a,t}$, $\widehat{F}_{a,t}$ and $\widehat{M}_{a,t}$ (see the color bar on the left) and, by excluding from the feasible space the points below the plane, the frequency never falls below the prescribed minimum level.

In the case of BCS, the maximum IFD depends on the weighted model parameters of the system that experiences the contingency and the supporting systems. Thus the 3D hyperplane becomes a 7D hyperplane (or more), and the IFD constraint is included as follows:
\begin{equation}\label{eq:nadir_bi}
\footnotesize
    \begin{split}
        \widetilde{F}_{a,t} \geq \sum_{b\in\mathcal{A}_a} & \big(A^{\textsc{R}^g}_{a,b} \widetilde{R}^g_{b,t} + A^{\textsc{M}}_{a,b} \widehat{M}_{b,t} + A^{\textsc{F}}_{a,b} \widetilde{F}^g_{b,t} + A^{\textsc{R}^c}_{a,b} \widetilde{R}^c_{a,b,t}\big) \\
        &+ A^{\textsc{R}}_a \widetilde{R}^g_{a,t} + A^{\textsc{M}}_a \widehat{M}_{a,t} + A^0_a \quad\quad \forall a,\,\forall t
    \end{split}
\end{equation}
with $\mathcal{A}_a$ the set of asynchronous areas connected to area $a$, and $A^{\textsc{R}^g}_{a,b}$, $A^{\textsc{M}}_{a,b}$, $A^{\textsc{F}}_{a,b}$ and $A^{\textsc{R}^c}_{a,b}$ the corresponding coefficients of the hyperplane. The three 3D plots on the right, in \figurename~\ref{fig:hyperplane}, provide a representation of the 7D hyperplane (depicted in red) calculated for two interconnected areas. These plots are obtained by fixing the aggregated system parameters of one area (in this case area b) and by varying the sum of converter droop gains ($\widetilde{R}^c_{a,b}$) between the two areas. Similar to the left plot, only the points above the hyperplane satisfy the IFD constraints.

This data-driven approach, which consists in (i) calculating the IFD for a large number of possible combinations of system parameters and (ii) finding the points on the boundary to be approximated by the hyperplane, significantly increases the computational speed of the problem while keeping the relative error small (average relative error 0.04).

\vspace{-0.5em}
\subsection{Post-contingency Constraints}
With droop frequency controllers, the response of generators and HVDC converters is proportional to the ratio between their droop coefficient and the total droop in the system. To ensure that generators committed to provide frequency services and HVDC converters with SPC have enough capacity to support the system, some capacity must be reserved. At the same time, domestic reserves must be enough to guarantee system security in case of HVDC link outages. The following constraints model the response of generator and HVDC converters and ensure enough capacity reservation:

\begin{subequations}
\small\vspace{-1em}
\begin{alignat}{2}
    & g^s_{i,t} = v^g_{i,t} \tfrac{K^g_i\overline{P}_i}{R^g_i\widehat{R}_{(a|i\in\mathcal{G}_a),t}} \Delta P^e_{(a|i\in\mathcal{G}_a)} \quad\quad && \forall i,\,\forall t \label{eq:gres}\\
    & g_{i,t} + g^s_{i,t} \leq u_{i,t} \overline{P}_i \quad\quad && \forall i,\,\forall t \label{eq:gres_lim}\\
    & \textstyle\sum_{i\in\mathcal{G}_a}g^s_{i,t} \geq \Delta P^{\textsc{dc}} \quad\quad && \forall t \label{eq:HVDCfault}\\
    & p^s_{k,t} = v^c_{(a|k\in\mathcal{L}^{\textsc{dc}}_a),\sim,t} \tfrac{K^c_k\overline{P}^{\textsc{dc}}_{k}}{R^c_k\widehat{R}_{(a|k\in\mathcal{L}^{\textsc{dc}}_a),t}} \Delta P^e_{(a|k\in\mathcal{L}^{\textsc{dc}}_a)} \quad && \forall k,\,\forall t \label{eq:fres}\\
    & -\overline{P}^{\textsc{dc}}_{k} \leq p^{\textsc{dc}}_{k,t} + p^{s}_{k,t}\leq \overline{P}^{\textsc{dc}}_{k} \qquad\qquad\qquad\quad \forall k\in && \mathcal{L}^{\rm vsc}, \forall t \label{eq:fres_limVSC}\\
    & -(1-u^{\rm lcc}_k)\overline{P}^{\textsc{dc}}_{k} \leq p^{\textsc{dc}}_{k,t} + p^{s}_{k,t}\leq u^{\rm lcc}_k\overline{P}^{\textsc{dc}}_{k} \quad\, \forall k\in && \mathcal{L}^{\rm lcc}, \forall t \label{eq:fres_limLCC}
\end{alignat}
\end{subequations}
with $g^s_{i,t}$ the response of generator $i$ (equal to the reserved capacity), $\Delta P^{\textsc{dc}}$ the maximum power deviation following an HVDC link outage (equal to the full capacity of monopole links or to half of the capacity for bipole links), and  $p^s_{k,t}$ the response of the HVDC converter $k$ (and the reserved capacity). Constraints \eqref{eq:HVDCfault} guarantees that, in case of HVDC line outage, there are enough frequency reserves to guarantee system security in each area. Similar to constraint \eqref{eq:dc_limLCC}, constraint \eqref{eq:fres_limLCC} ensures that the polarity of LCC converters is not inverted to provide frequency support. Constraints \eqref{eq:gres} and \eqref{eq:fres} are non-linear because $\widehat{R}_{a,t}$ appears in the denominator and the fraction is then multiplied by $v^g_{i,t}$ or $v^c_{a,\sim,t}$. In the following, two techniques to linearize these expressions are presented. 

\vspace{-0.5em}
\subsection{Relaxation techniques}
In order to linearize contraints \eqref{eq:gres} and \eqref{eq:fres}, the first step is to introduce a new variable $\widehat{S}_{a,t}$ such that:
\begin{equation} \label{Eq:RSbil}
\small
\widehat{R}_{a,t}\widehat{S}_{a,t} = 1 \quad\quad \forall a,\,\forall t
\end{equation}
This constraint is bilinear and, in order to linearize it, the McCormick relaxation is applied \cite{UCP_2}. Unless the bounds of the variables are quite small, this relaxation is not tight. To improve it, a piecewise version of the relaxation is applied, as suggested in \cite{UCP_3}. Constraint \eqref{Eq:RSbil} is then substituted by:

\begin{subequations}
\footnotesize\vspace{-1em}
\begin{alignat}{2}
    & \textstyle\sum_{x,y} \big( \underline{S}^a_yR^a_{x,y,t} + \underline{R}^a_xS^a_{x,y,t} - \underline{S}^a_y\underline{R}^a_xw^a_{x,y,t} \big) \leq 1 \quad\quad && \forall a,\,\forall t \label{eq:mcc_1}\\
    & \textstyle\sum_{x,y} \big( \overline{S}^a_yR^a_{x,y,t} + \overline{R}^a_xS^a_{x,y,t} - \overline{S}^a_y\overline{R}^a_xw^a_{x,y,t} \big) \leq 1 \quad\quad && \forall a,\,\forall t \\
    & \textstyle\sum_{x,y} \big( \overline{S}^a_yR^a_{x,y,t} + \underline{R}^a_xS^a_{x,y,t} - \overline{S}^a_y\underline{R}^a_xw^a_{x,y,t} \big) \geq 1 \quad\quad && \forall a,\,\forall t \\
    & \textstyle\sum_{x,y} \big( \underline{S}^a_yR^a_{x,y,t} + \overline{R}^a_xS^a_{x,y,t} - \underline{S}^a_y\overline{R}^a_xw^a_{x,y,t} \big) \geq 1 \quad\quad && \forall a,\,\forall t \\
    & \widehat{R}_{a,t} = \textstyle\sum_{x,y}R^a_{x,y,t}, \quad \widehat{S}_{a,t} = \textstyle\sum_{x,y}S^a_{x,y,t} \quad\quad && \forall a,\,\forall t \\
    & w^a_{x,y,t}\underline{R}^a_x \leq R^a_{x,y,t} \leq w^a_{x,y,t}\overline{R}^a_x \quad\quad\quad\quad\quad\quad \forall x,\,\forall y,\,&&\forall a,\,\forall t \\
    & w^a_{x,y,t}\underline{S}^a_y \leq S^a_{x,y,t} \leq w^a_{x,y,t}\overline{S}^a_y \quad\quad\quad\quad\quad\quad\,\, \forall x,\,\forall y,\,&&\forall a,\,\forall t \\
    & \textstyle\sum_{x,y}w^a_{x,y,t} = 1\quad\quad && \forall a,\,\forall t \label{eq:mcc_end}
\end{alignat}
\end{subequations}

where $R^a_{x,y,t}$ and $R^a_{x,y,t}$ are new continuous variables (for the sake of space the index $a$ is written as superscript), $w^a_{x,y,t}$ new binary variables and $\underline{R}^a_x$, $\overline{R}^a_x$, $\underline{S}^a_y$ and $\overline{S}^a_y$ the lower and upper bounds defining the segments in which  $\widehat{R}_{a,t}$ and $\widehat{S}_{a,t}$ are divided. The subscript $x$ refers to the $x$-th segment of $\widehat{R}_{a,t}$ and $y$ to the $y$-th segment of $\widehat{S}_{a,t}$. Given that $\widehat{S}_{a,t}$ is the reciprocal of $\widehat{R}_{a,t}$, the division of $\widehat{S}_{a,t}$ in segments is linked to the division of $\widehat{R}_{a,t}$, meaning that high accuracy can be obtained with few segments. 

Similar to \eqref{Eq:RSbil}, the terms $v^g_{i,t}\widehat{S}_{a,t}$ and $v^c_{a,\sim,t}\widehat{S}_{a,t}$, which appear in constraints \eqref{eq:gres} and \eqref{eq:fres} after the introduction of the new variable, must be linearized. In this case, these terms are the product of a binary and a continuous variable, thus the \textit{bigM} method can be applied \cite{UCP_5}. Constraints \eqref{eq:gres} and \eqref{eq:fres} are then replaced by:

\begin{subequations}
\footnotesize\vspace{-1em}
\begin{alignat}{2}
    & g^s_{i,t} = \tfrac{K^g_i\overline{P}_i}{R^g_i} \widehat{VS}^g_{i,(a|i\in\mathcal{G}_a),t} \Delta P^e_{(a|i\in\mathcal{G}_a)} \quad\quad && \forall i,\,\forall t \label{eq:bigM_1}\\
    & p^s_{k,t} = \tfrac{K^c_k\overline{P}^{\textsc{dc}}_k}{R^c_k} \widehat{VS}^c_{k,(a|k\in\mathcal{L}^{\textsc{dc}}_a),\sim,t} \Delta P^e_{(a|i\in\mathcal{G}_a)} \quad\quad && \forall k,\,\forall t \\
    & 0 \leq \widehat{VS}^g_{i,(a|i\in\mathcal{G}_a),t} \leq v^g_{i,t} M \quad\quad && \forall i,\,\forall t \label{eq:Mbound}\\
    & \widehat{S}_{a,t} - (1-v^g_{i,t})M \leq \widehat{VS}^g_{i,(a|i\in\mathcal{G}_a),t} \leq \widehat{S}_{a,t} \quad\quad && \forall i,\,\forall t \\
    & 0 \leq \widehat{VS}^c_{k,(a|k\in\mathcal{L}^{\textsc{dc}}_a),\sim,t} \leq v^c_{a,\sim,t} M \quad\quad && \forall i,\,\forall t \\
    & \widehat{S}_{a,t} - (1-v^c_{a,\sim,t})M \leq \widehat{VS}^c_{k,(a|k\in\mathcal{L}^{\textsc{dc}}_a),\sim,t} \leq \widehat{S}_{a,t} \quad\quad && \forall i,\,\forall t \label{eq:bigM_end}
\end{alignat}
\end{subequations}
with $\widehat{VS}^g_{i,(a|i\in\mathcal{G}_a),t}$ and $\widehat{VS}^c_{k,(a|k\in\mathcal{L}^{\textsc{dc}}_a),\sim,t}$ the new continuous variables. As pointed out in \cite{UCP_6}, the selection of the right $M$ is a delicate process: with too small values, constraint \eqref{eq:Mbound} might be binding even when $v^g_{i,t}$ is equal to one. On the other hand, too large values might cause numerical inaccuracies when solving the problem. In this case, since $\widehat{S}_{a,t}$ is the reciprocal of $\widehat{R}_{a,t}$, and $\widehat{R}_{a,t} \geq 1$, it is possible to set $M=10$ without incurring in any of these problems.

\vspace{-0.5em}
\subsection{Optimization Problem}
The UCP with frequency constraints, HVDC SPC and generator frequency support is formulated as:
{\small
\begin{alignat}{2}
    & \text{min} \enspace && \eqref{eq:obj} \nonumber\\
    & \text{s.t.} \enspace \enspace && \eqref{eq:gen1} - \eqref{eq:shutdown},\,\,\eqref{eq:ac_lim} - \eqref{eq:dc_limLCC},\,\,\eqref{eq:bal} \nonumber\\
    & && \eqref{eq:v_gen},\,\,\eqref{eq:R_g} - \eqref{eq:M},\,\,\eqref{eq:rocof},\,\,\eqref{eq:ssfd} \label{eq:opt_problem}\\
    & && \eqref{eq:v_uni},\,\,\eqref{eq:nadir_uni}\,\,\text{or}\,\,\eqref{eq:v_bi},\,\,\eqref{eq:nadir_bi}  \nonumber\\
    & && \eqref{eq:mcc_1} - \eqref{eq:mcc_end},\,\,\eqref{eq:bigM_1} - \eqref{eq:bigM_end} \nonumber\\
    & && \eqref{eq:gres_lim},\,\,\eqref{eq:HVDCfault},\,\,\eqref{eq:fres_limVSC},\,\,\eqref{eq:fres_limLCC}  \nonumber
\end{alignat}%
}%
Constraints \eqref{eq:v_uni} and \eqref{eq:nadir_uni} are enforced in case of unilateral exchange of reserves; for the bilateral exchange, instead, constraints \eqref{eq:v_bi} and \eqref{eq:nadir_bi} are introduced. In case of different control schemes applied between different zones, a combination of these constraints is included.

The energy prices can be calculated as suggested in \cite{UCP_7}: first the solution of the MILP problem is calculated, then the integer variables are relaxed to continuous variables and fixed to the obtained solution, and the problem (now a LP problem) is solved again. In this way, Lagrangian multipliers can be obtained. Energy prices are then calculated as the derivative of the Lagrangian function of the relaxed problem ($\mathcal{L}^*$) with respect to a marginal increase $\delta$ in the demand:
\begin{equation}\label{eq:prices}\small
     \text{EP}_n = \frac{\partial \mathcal{L}^*}{\partial \delta} = \lambda_n - \textstyle\sum_{d\in\mathcal{D}_n} \rho_d
\end{equation}
with $\lambda_n$ the Lagrangian multiplier associated with the nodal balance equation (bus $n$) and $\rho_d$ the multiplier associated with the upper bound on load shedding (load $d$). 


For the numerical examples in the next section, the reserves are procured on a daily basis and the energy market is cleared for a 24-hour time period. This means that, if a generator (or HVDC converter) commits to provide frequency support, the capacity is reserved for the whole day. The reasoning behind this is that, in Europe, most of the TSOs procure reserves in D-2 or D-1 for the entire day of operation, and the energy market is cleared once for every hour of the following day. However, the presented formulation is general and it also allows for hourly procurement if required. Also, the unit commitment problem could be solved for longer time windows. In case the size of the UCP would make it intractable, decomposition techniques such as the one in \cite{Hua2018} could be applied to decompose the problem by time instances. In this paper, the problem is solved in a centralized fashion, as we look at it from a total system perspective. Our aim is to deliver a tool that explicitly considers how the existing capabilities of HVDC lines allow for the sharing of reserves between asynchronous areas, and is able to quantify the reduction in the total system costs. In a European market setting, performing a spatial decomposition of our problem, by following e.g. the technique proposed in \cite{Hua2018} and standard decomposition techniques, we can obtain a market tool that can achieve similar results, and can be used by the TSOs in each respective area, similar to the Unit Commitment that currently takes place. Finally, the problem formulation is presented in a deterministic setup; however, the problem formulation could also be solved in a probabilistic fashion.

\section{Numerical Results}\label{sec:5}
In this section, two numerical examples are presented. The first example is provided to study the impact of exchanging frequency services via HVDC on the market outcome and the market participants. A simple 2-area system is used for the simulation: the two asynchronous areas are connected by a single VSC-HVDC link and the internal AC networks are omitted to enhance the readability of the results. The second example, instead, is used to prove the validity of the model in case of more asynchronous areas interconnected, and to study the scalability of the model considering internal AC networks, different HVDC technologies, and a larger number of market participants. All the simulations are run using YALMIP \cite{yalmip} and Gurobi \cite{gurobi}.

\vspace{-0.5em}
\subsection{2-area System}

The two areas are approx. of the same size: the installed generating capacity is 13 and 11 GW respectively (35 and 30 generating units), with a peak demand of 9 and 6 GW. The synchronous generation mix includes nuclear, natural gas, oil, hard coal, lignite and biomass steam units; the maximum RES penetration in the two areas is 47\% and 55\%. Two nuclear power plants are the dimensioning incidents in the two areas, respectively 600 MW and 550 MW. The transfer capacity of the VSC-HVDC link (bipole) is 500 MW. The demand, wind and solar time series correspond to the actual profiles of East and West Denmark during July 2019. The dynamic parameters of generators are based on \cite{FMOD_2}, while the economic parameters are taken from \cite{TESTCASE}. The system data is available in \cite{github}.

To assess the benefit of the SPC implementation, four simulations are executed: i) ``\textit{no lim}'', the frequency constraints are omitted in the model and the reserve procurement is done by enforcing a minimum reserve requirement, ii) ``\textit{no SPC}'', the reserve procurement is done in accordance to the frequency constraints, iii) ``\textit{unilateral}'' and iv) ``\textit{bilateral}'', HVDC lines are involved in the exchange of frequency services, first with the unilateral scheme and then with the bilateral.

By looking at \figurename~\ref{fig:frequency}, it is clear that frequency constraints are violated if they are not included in the model (blue lines). This does not happen in the other simulations, showing that procuring reserves only on a cost basis without considering the response of the system is not enough to ensure its stability. 
Moreover, from \figurename~\ref{fig:reserves}, it is clear that the exchange of frequency reserves through HVDC can help reducing the costs of reserve procurement while maintaining the system N-1 secure. In terms of system costs, the total costs decrease by 1\% in the ``\textit{unilateral}'' case and by 1.4\% in the ``\textit{bilateral}''. The comparison is made with respect to ``\textit{no SPC}'' (total system cost 69.95 M\euro), as in the ``\textit{no lim}'' case (total system cost 68.16 M\euro) frequency constraints are not included (and are also violated) and thus the comparison would not be fair. The cost reduction is more pronounced if we only look at the reserve procurement costs, which decrease by 8.5\% in the ``\textit{unilateral}'' case and by 10.8\% in the ``\textit{bilateral}'' (in the ``\textit{no SPC}'' case the cost is 3.9 M\euro).

\definecolor{green1_plot}{rgb}{0.1608,0.4196,0.0745}%
\definecolor{green2_plot}{rgb}{0.4667,0.6745,0.1882}%
\definecolor{yellow_plot}{rgb}{0.9294,0.6902,0.1176}%
\definecolor{blue_plot}{rgb}{0,0.4392,0.7373}%
\definecolor{gray_plot}{rgb}{0.49412,0.49412,0.49412}%

\begin{figure}[!t]
    \vspace{-0.6em}
        \begin{tikzpicture}
            \begin{axis}[%
                width=0.205\textwidth,
                height=0.11\textheight,
                at={(0in,0in)},
                scale only axis,
                scale only axis,
                xmin=0,
                xmax=362,
                xticklabels={},
                ymin=0.30,
                ymax=0.92,
                yticklabel style={font=\footnotesize},
                ylabel style={font=\footnotesize},
                ylabel={RoCoF (Hz/s)},
                axis background/.style={fill=white},
                title style={font=\footnotesize,yshift=-0.25cm},
                title={Area 1},
                legend columns=4,
                legend style={at={(1.5in,-0.21in)}, anchor=south, legend cell align=left, align=left, draw=none, line width=0.2, font=\scriptsize},
                /tikz/every even column/.append style={column sep=0.3cm},
                every axis legend/.append style={column sep=0.3em},
                ]
                
                \addplot[draw=red,line width=0.5,dashed,forget plot] table[]{Plots/Data/rocof_limit.dat};
                
                \addplot[draw=blue_plot,line width=0.6,on layer=f1] table[]{Plots/Data/rocof_a1_nolim.dat};
                
                \addplot[draw=yellow_plot,line width=0.6,on layer=f1] table[]{Plots/Data/rocof_a1_noepc.dat};
                
                \addplot[draw=green1_plot,line width=0.6,on layer=f1] table[]{Plots/Data/rocof_a1_unil.dat};
                
                \addplot[draw=green2_plot,line width=0.6,dashed,on layer=f1] table[]{Plots/Data/rocof_a1_bil.dat};
                \legend{\textit{no lim},\textit{no SPC},\textit{unilateral},\textit{bilateral}}
            \end{axis}
            \begin{axis}[%
                width=0.205\textwidth,
                height=0.11\textheight,
                at={(1.55in,0in)},
                scale only axis,
                scale only axis,
                xmin=0,
                xmax=362,
                xticklabels={},
                ymin=0.3,
                ymax=0.92,
                yticklabels={},
                axis background/.style={fill=white},
                title style={font=\footnotesize,yshift=-0.25cm},
                title={Area 2},
                ]
                \addplot [draw=gray!60, line width=0.8pt, forget plot] table[row sep=crcr]{%
                43	0.3	\\
                43  0.9	\\
                };
                \addplot[draw=red,line width=0.5,dashed,forget plot] table[]{Plots/Data/rocof_limit.dat};
                
                \addplot[draw=blue_plot,line width=0.6,on layer=f1] table[]{Plots/Data/rocof_a2_nolim.dat};
                
                \addplot[draw=yellow_plot,line width=0.6,on layer=f1] table[]{Plots/Data/rocof_a2_noepc.dat};
                
                \addplot[draw=green1_plot,line width=0.6,on layer=f1] table[]{Plots/Data/rocof_a2_unil.dat};
                
                \addplot[draw=green2_plot,line width=0.6,dashed,on layer=f1] table[]{Plots/Data/rocof_a2_bil.dat};
            \end{axis}
    \end{tikzpicture}
    \begin{tikzpicture}
            \begin{axis}[%
                width=0.205\textwidth,
                height=0.11\textheight,
                at={(0in,0in)},
                scale only axis,
                scale only axis,
                xmin=0,
                xmax=362,
                xticklabel style={font=\footnotesize},
                xlabel style={font=\footnotesize},
                xlabel={Hours (h)},
                ymin=0.45,
                ymax=0.9,
                ytick = {0.5,0.6,0.7,0.8},
                yticklabel style={font=\footnotesize},
                ylabel style={font=\footnotesize},
                ylabel={IFD (Hz)},
                axis background/.style={fill=white},
                ]
                
                \addplot[draw=red,line width=0.5,dashed] table[]{Plots/Data/nadir_limit.dat};
                
                \addplot[draw=blue_plot,line width=0.6,on layer=f1] table[]{Plots/Data/nadir_a1_nolim.dat};
                
                \addplot[draw=yellow_plot,line width=0.6,on layer=f1] table[]{Plots/Data/nadir_a1_noepc.dat};
                
                \addplot[draw=green1_plot,line width=0.6,on layer=f1] table[]{Plots/Data/nadir_a1_unil.dat};
                
                \addplot[draw=green2_plot,line width=0.6,dashed,on layer=f1] table[]{Plots/Data/nadir_a1_bil.dat};
            \end{axis}
            \begin{axis}[%
                width=0.205\textwidth,
                height=0.11\textheight,
                at={(1.55in,0in)},
                scale only axis,
                scale only axis,
                xmin=0,
                xmax=362,
                xticklabel style={font=\footnotesize},
                xlabel style={font=\footnotesize},
                xlabel={Hours (h)},
                ymin=0.45,
                ymax=0.9,
                ytick = {0.5,0.6,0.7,0.8},
                yticklabels={},
                axis background/.style={fill=white},
                ]
                \addplot [draw=gray!60, line width=0.8pt, forget plot] table[row sep=crcr]{%
                43	0.45	\\
                43  0.9	\\
                };
                \addplot[draw=red,line width=0.5,dashed] table[]{Plots/Data/nadir_limit.dat};
                
                \addplot[draw=blue_plot,line width=0.6,on layer=f1] table[]{Plots/Data/nadir_a2_nolim.dat};
                
                \addplot[draw=yellow_plot,line width=0.6,on layer=f1] table[]{Plots/Data/nadir_a2_noepc.dat};
                
                \addplot[draw=green1_plot,line width=0.6,on layer=f1] table[]{Plots/Data/nadir_a2_unil.dat};
                
                \addplot[draw=green2_plot,line width=0.6,dashed,on layer=f1] table[]{Plots/Data/nadir_a2_bil.dat};
            \end{axis}
    \end{tikzpicture}
    \vspace{-2em}
    \caption{Frequency metrics for the four simulations. The red dashed lines are the limits (0.625 Hz/s for the RoCoF and 0.7 Hz for the IFD), while the gray vertical ones highlight the time instance used for validation.}
    \label{fig:frequency}
\end{figure}
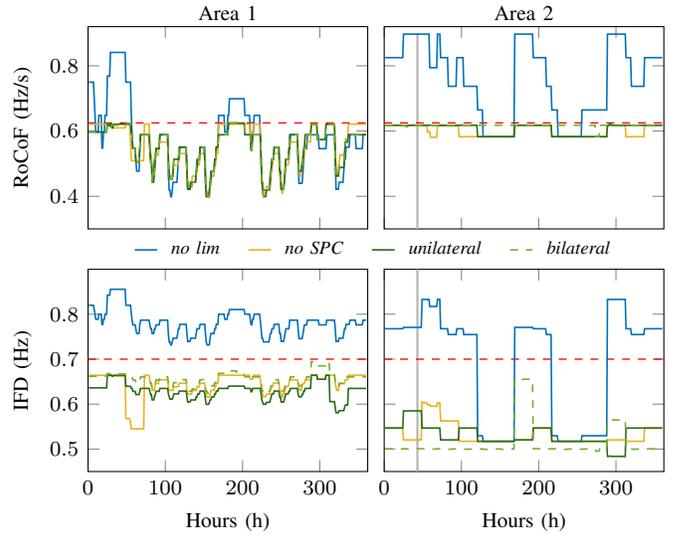
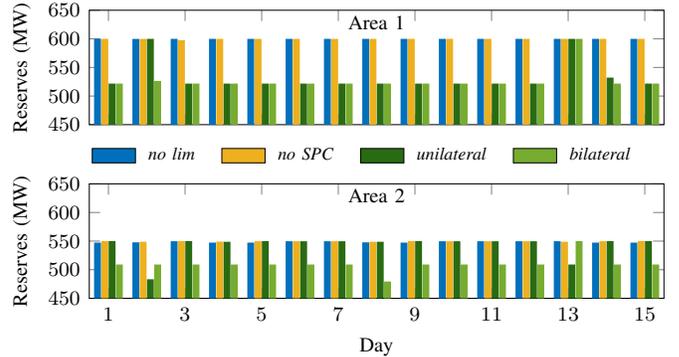
\begin{figure}[!t]
    \vspace{-0.6em}
    \resizebox{0.49\textwidth}{!}{%
        \begin{tikzpicture}
            \begin{axis}[%
                width=0.44\textwidth,
                height=0.065\textheight,
                at={(0in,0in)},
                scale only axis,
                bar shift=0,
                xmin=0.5,
                xmax=15.5,
                xtick={1,3,5,7,9,11,13,15},
                xticklabels={},
                xlabel style={font=\footnotesize},
                ymin=450,
                ymax=650,
                yticklabel style={font=\footnotesize},
                ylabel style={font=\footnotesize},
                ylabel={Reserves (MW)},
                axis background/.style={fill=white},
                title style={font=\footnotesize,yshift=-0.6cm},
                title={Area 1},
                legend columns=4,
                legend style={at={(1.5in,-0.27in)}, anchor=south, legend cell align=left, align=left, draw=none, line width=0.2, font=\scriptsize},
                /tikz/every even column/.append style={column sep=0.3cm},
                every axis legend/.append style={column sep=0.3em},
                ]
                
                \addplot[ybar, bar width=0.18, fill=blue_plot, draw=none, area legend] table[x index=0,y index=4]{Plots/Data/reserves.dat};
                \addplot[ybar, bar width=0.18, fill=yellow_plot, draw=none, area legend] table[x index=1,y index=5]{Plots/Data/reserves.dat};
                \addplot[ybar, bar width=0.18, fill=green1_plot, draw=none, area legend] table[x index=2,y index=6]{Plots/Data/reserves.dat};
                \addplot[ybar, bar width=0.18, fill=green2_plot, draw=none, area legend] table[x index=3,y index=7]{Plots/Data/reserves.dat};
                \legend{\textit{no lim},\textit{no SPC},\textit{unilateral},\textit{bilateral}}
            \end{axis}
            \begin{axis}[%
                width=0.44\textwidth,
                height=0.065\textheight,
                at={(0in,-0.95in)},
                scale only axis,
                bar shift=0,
                xmin=0.5,
                xmax=15.5,
                xtick={1,3,5,7,9,11,13,15},
                xticklabel style={font=\footnotesize},
                xlabel={Day},
                xlabel style={font=\footnotesize},
                ymin=450,
                ymax=650,
                yticklabel style={font=\footnotesize},
                ylabel style={font=\footnotesize},
                ylabel={Reserves (MW)},
                axis background/.style={fill=white},
                title style={font=\footnotesize,yshift=-0.6cm},
                title={Area 2},
                ]
                \addplot[ybar, bar width=0.18, fill=blue_plot, draw=none, area legend] table[x index=0,y index=8]{Plots/Data/reserves.dat};
                \addplot[ybar, bar width=0.18, fill=yellow_plot, draw=none, area legend] table[x index=1,y index=9]{Plots/Data/reserves.dat};
                \addplot[ybar, bar width=0.18, fill=green1_plot, draw=none, area legend] table[x index=2,y index=10]{Plots/Data/reserves.dat};
                \addplot[ybar, bar width=0.18, fill=green2_plot, draw=none, area legend] table[x index=3,y index=11]{Plots/Data/reserves.dat};
            \end{axis}
        \end{tikzpicture}
    }%
    \vspace{-1.2em}
    \caption{Generation capacity reserved during each day of the simulations.}
    \label{fig:reserves}
    \vspace{-1.2em}
\end{figure}

The impact of the activation of HVDC SPC on the market outcome is provided in Table~\ref{tab:market}. The amount of generator capacity that is not reserved (because of SPC), results in more capacity available in the energy market. Since generators must be dispatched to provide support, this means that local production increases in the importing area (without affecting the energy price), while exports decrease in the other area. This means more revenues for the generators in the importing area and less revenues for the ones in the exporting area. At the same time, the increased generation results in higher energy production and lower reserve procurement costs in area 1. 

Looking at load payments, these increase in area 2 while decreasing in area 1. The explanation for this is that the hours with congestion decrease (respectively 283, 222, 163 and 188 hours in the four simulations) meaning that consumers in the two areas are subject to the same price for more hours. Given that area 1 is the ``high price'' area and area 2 the ``low price'' one, the result is lower payments from the loads in area 1 and higher payments from the loads in area 2.

\begin{table}[!t]
\vspace{-0.5em}
    \caption{Impact of HVDC SPC on the market outcome.}
    \label{tab:market}
    \small
    \centering
    \resizebox{0.48\textwidth}{!}{%
        \begin{tabular}{lcrrrr}
        \hline
                       & 		 & \textit{no lim} & \textit{no SPC}\footnotemark[1] & \textit{unilateral} & \textit{bilateral} \Tstrut\Bstrut\\ 
        \hline
        \multirow{2}{*}{Reserve Cost}   & \textsc{area} 1    & -2.2\%    & (1.77)    & -12.3\%   & -14.21\% \Tstrut\\
                                        & \textsc{area} 2    & -11.6\%   & (2.15)    & -5.3\%    & {-7.9\% } \\
        \multirow{2}{*}{Energy Cost}    & \textsc{area} 1    & +2.6\%    & (35.86)   & +1.5\%    & +1.6\% \Tstrut\\ 
                                        & \textsc{area} 2    & -8.1\%    & (30.17)   & -3.1\%    & {-3.8\% } \\
        \multirow{2}{*}{Total Cost}     & \textsc{area} 1    & +2.3\%    & (37.63)   & +0.9\%    & +0.8\% \Tstrut\\
                                        & \textsc{area} 2    & -8.3\%    & (32.33)   & -3.2\%    & -4.1\% \Bstrut\\ 
        \hline
        \multirow{2}{*}{Gen. Rev.}      & \textsc{area} 1    & +13.1\%   & (28.89)   & +5.7\%    & +5.9\% \Tstrut\\
                                        & \textsc{area} 2    & +27.1\%   & (15.67)   & -3.4\%    & {+0.2\% } \\
        \multirow{2}{*}{Load Pymt.}     & \textsc{area} 1	& -5.0\%      & (71.16)   & -1.5\%    & -1.7\% \Tstrut\\
                                        & \textsc{area} 2    & -7.5\%    & (42.85)   & +0.8\%    & {- } \\
        HVDC Rev.                   & -         & -54.0\%     & (0.23)    & +21.0\% & +9.0\% \Tstrut\Bstrut\\         
        \hline
    \end{tabular}
    }%
\end{table}

\begin{table}[!t]
\vspace{-1em}
    \caption{Comparison of the IFD obtained with the analytical expressions and the dynamic simulations.}
    \label{tab:validation}
    \small
    \centering
    \resizebox{0.48\textwidth}{!}{%
        \begin{tabular}{lcrrrr}
        \hline
                       & 		 & \textit{no lim} & \textit{no SPC} & \textit{unilateral} & \textit{bilateral} \Tstrut\Bstrut\\ 
        \hline
        Analytical  & \textsc{area} 1    & -        & -         & 0.1550    & 0.1496 \Tstrut\\
        expression  & \textsc{area} 2    & 0.7707   & 0.5204    & 0.5849    & 0.5007 \Bstrut\\
        Dynamic     & \textsc{area} 1    & -        & -         & 0.1261    & 0.1001 \Tstrut\\
        simulation  & \textsc{area} 2    & 0.7706   & 0.5204    & 0.5896    & 0.4875 \Bstrut\\
        \hline
    \end{tabular}
    }%
\end{table}

Concerning the HVDC line owner, one could expect lower congestion rents due to the reservation of a portion of the capacity. However, the lower capacity results in higher price differences (respectively 0.58, 1.28, 2.04 and 1.81 \euro/MWh in average) and, in turn, in higher congestion rent.

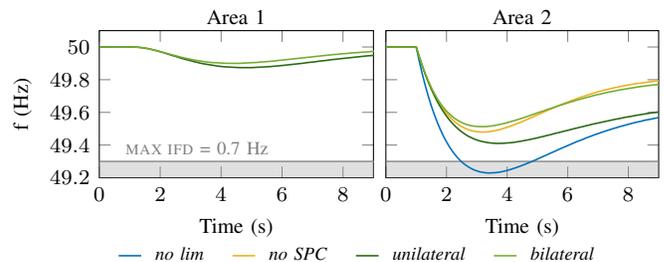
\begin{figure}[!t]
    \vspace{-0.6em}
        \begin{tikzpicture}
            \begin{axis}[%
                width=0.201\textwidth,
                height=0.08\textheight,
                at={(0in,0in)},
                scale only axis,
                scale only axis,
                xmin=0,
                xmax=9,
                xticklabel style={font=\footnotesize},
                xlabel style={font=\footnotesize},
                xlabel={Time (s)},
                ymin=49.2,
                ymax=50.1,
                yticklabel style={font=\footnotesize},
                ylabel style={font=\footnotesize},
                ylabel={f (Hz)},
                axis background/.style={fill=white},
                title style={font=\footnotesize,yshift=-0.25cm},
                title={Area 1}
                ]
                \addplot[draw=gray_plot,line width=0.5, forget plot] table[x index=0,y index=7]{Plots/Data/data_validation.dat};
                \addplot[area legend, draw=none, fill=white!60!black, fill opacity=0.3, forget plot]
                table[row sep=crcr] {%
                x   y\\
                0   49\\
                0   49.3\\
                10  49.3\\
                10  49\\
                }--cycle;
                \addplot[forget plot] coordinates {(3.2, 49.48)} node[below] {\textcolor{gray_plot}{\scriptsize{\textsc{max ifd} = 0.7 Hz}}};
                \addplot[draw=green1_plot,line width=0.6] table[x index=0,y index=1]{Plots/Data/data_validation.dat};
                \addplot[draw=green2_plot,line width=0.6] table[x index=0,y index=2]{Plots/Data/data_validation.dat};
            \end{axis}
            \begin{axis}[%
                width=0.20\textwidth,
                height=0.08\textheight,
                at={(1.5in,0in)},
                scale only axis,
                scale only axis,
                xmin=0,
                xmax=9,
                xticklabel style={font=\footnotesize},
                xlabel style={font=\footnotesize},
                xlabel={Time (s)},
                ymin=49.2,
                ymax=50.1,
                yticklabels={},
                axis background/.style={fill=white},
                title style={font=\footnotesize,yshift=-0.25cm},
                title={Area 2},
                legend columns=4,
                legend style={at={(-0.1,-0.4)}, anchor=north, draw=none, line width=0.2, font=\scriptsize},
                /tikz/every even column/.append style={column sep=0.3cm},
                every axis legend/.append style={column sep=0.3em},
                ]
                \addplot[draw=gray_plot,line width=0.5, forget plot] table[x index=0,y index=7]{Plots/Data/data_validation.dat};
                \addplot[area legend, draw=none, fill=white!60!black, fill opacity=0.3, forget plot]
                table[row sep=crcr] {%
                x   y\\
                0   49\\
                0   49.3\\
                10  49.3\\
                10  49\\
                }--cycle;
                \addplot[draw=blue_plot,line width=0.6] table[x index=0,y index=3]{Plots/Data/data_validation.dat};
                \addplot[draw=yellow_plot,line width=0.6] table[x index=0,y index=4]{Plots/Data/data_validation.dat};
                \addplot[draw=green1_plot,line width=0.6] table[x index=0,y index=5]{Plots/Data/data_validation.dat};
                \addplot[draw=green2_plot,line width=0.6] table[x index=0,y index=6]{Plots/Data/data_validation.dat};
                \legend{\textit{no lim},\textit{no SPC},\textit{unilateral},\textit{bilateral}}
            \end{axis}
    \end{tikzpicture}
    \vspace{-1em}
    \caption{IFD following the dimensioning incident in area 2. The left graph shows the IFD in area 1 due to the activation of HVDC SPC.}
    \label{fig:validation}
    \vspace{-0.5em}
\end{figure}

\begin{figure}[!t]
        \begin{tikzpicture}
            \begin{axis}[%
                width=0.201\textwidth,
                height=0.08\textheight,
                at={(0in,0in)},
                scale only axis,
                scale only axis,
                xmin=0,
                xmax=9,
                xticklabel style={font=\footnotesize},
                xlabel style={font=\footnotesize},
                xlabel={Time (s)},
                ymin=49.2,
                ymax=50.8,
                yticklabel style={font=\footnotesize},
                ylabel style={font=\footnotesize},
                ylabel={f (Hz)},
                axis background/.style={fill=white},
                title style={font=\footnotesize,yshift=-0.25cm},
                title={Area 1}
                ]

                \addplot[area legend, draw=gray_plot, fill=white!60!black, fill opacity=0.3, forget plot]
                table[row sep=crcr] {%
                x   y\\
                0   49\\
                0   49.3\\
                10  49.3\\
                10  49\\
                }--cycle;
                \addplot[area legend, draw=gray_plot, fill=white!60!black, fill opacity=0.3, forget plot]
                table[row sep=crcr] {%
                x   y\\
                0   51\\
                0   50.7\\
                10  50.7\\
                10  51\\
                }--cycle;
                \addplot[forget plot] coordinates {(3.2, 49.6)} node[below] {\textcolor{gray_plot}{\scriptsize{\textsc{max ifd} = 0.7 Hz}}};
                \addplot[draw=green1_plot,line width=0.6] table[x index=0,y index=2]{Plots/Data/HVDCfault.dat};
                \addplot[draw=green2_plot,line width=0.6] table[x index=0,y index=3]{Plots/Data/HVDCfault.dat};
            \end{axis}
            \begin{axis}[%
                width=0.20\textwidth,
                height=0.08\textheight,
                at={(1.5in,0in)},
                scale only axis,
                scale only axis,
                xmin=0,
                xmax=9,
                xticklabel style={font=\footnotesize},
                xlabel style={font=\footnotesize},
                xlabel={Time (s)},
                ymin=49.2,
                ymax=50.8,
                yticklabels={},
                axis background/.style={fill=white},
                title style={font=\footnotesize,yshift=-0.25cm},
                title={Area 2},
                legend columns=4,
                legend style={at={(-0.1,-0.4)}, anchor=north, draw=none, line width=0.2, font=\scriptsize},
                /tikz/every even column/.append style={column sep=0.3cm},
                every axis legend/.append style={column sep=0.3em},
                ]
                \addplot[area legend, draw=gray_plot, fill=white!60!black, fill opacity=0.3, forget plot]
                table[row sep=crcr] {%
                x   y\\
                0   49\\
                0   49.3\\
                10  49.3\\
                10  49\\
                }--cycle;
                \addplot[area legend, draw=gray_plot, fill=white!60!black, fill opacity=0.3, forget plot]
                table[row sep=crcr] {%
                x   y\\
                0   51\\
                0   50.7\\
                10  50.7\\
                10  51\\
                }--cycle;
                \addplot[draw=green1_plot,line width=0.6] table[x index=0,y index=5]{Plots/Data/HVDCfault.dat};
                \addplot[draw=green2_plot,line width=0.6] table[x index=0,y index=6]{Plots/Data/HVDCfault.dat};
                \legend{\textit{unilateral},\textit{bilateral}}
            \end{axis}
    \end{tikzpicture}
    \vspace{-1em}
    \caption{IFD following the loss of one pole of the HVDC link. The left figure shows the under frequency event in area 1 (the importer), while the right figure shows the over frequency event in area 2 (the exporter).}
    \label{fig:HVDCfault}
\end{figure}
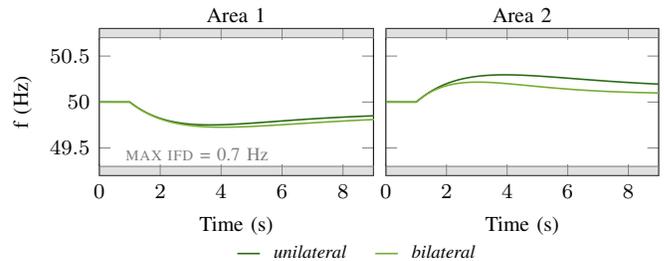

Table~\ref{tab:validation} compares the IFDs calculated with the analytical expressions to the ones obtained through dynamic simulation (\figurename~\ref{fig:validation}) for a specific time instance (hour 43). The two values match in all the cases, only in the ``\textit{bilateral}'' case the calculated IFD is a bit higher (this was expected due to the assumptions in Section \ref{sec:2}), meaning that the UC solution is more conservative. Finally, Fig. \ref{fig:HVDCfault} shows the frequency response of the two asynchronous interconnected areas following a HVDC-link loss. As it can be seen, the frequency deviation in both systems remains well below the IFD limit for both control schemes (``\textit{unilateral}'' and ``\textit{bilateral}''). This is because constraint \eqref{eq:HVDCfault} guarantees that there are enough reserves in each area and because the loss of the HVDC link (bipole) is less severe than the dimensioning incident (250 MW).

\footnotetext[1]{Used as a reference for the other simulation, the unit is million Euros.}

\vspace{-0.5em}
\subsection{3-area System}

A third area is now connected to area 1 and area 2: this area is significantly smaller than the others, with approx 3 GW of synchronous generation capacity (18 units), 0.7 GW of wind power and 2.2 GW of peak load. At the same time, the size of area 1 is doubled: 22.8 GW of synchronous generation (60 units), 4.3 GW of wind and 18.3 GW of peak load. Two VSC-HVDC links (bipole - 500 MW each) connect area 1 to area 2, one LCC-HVDC link (monopole - 200 MW) connects area 1 to area 3 and one VSC-HVDC link (monopole - 150 MW) connects area 2 to area 3. The three dimensioning incidents in the three areas are respectively 900 MW, 550 MW and 300 MW. In addition, the AC network of each area is now considered: area 1 and 2 consists of 6 electrical nodes, while area 3 has 3 nodes. The system data is available in \cite{github}.

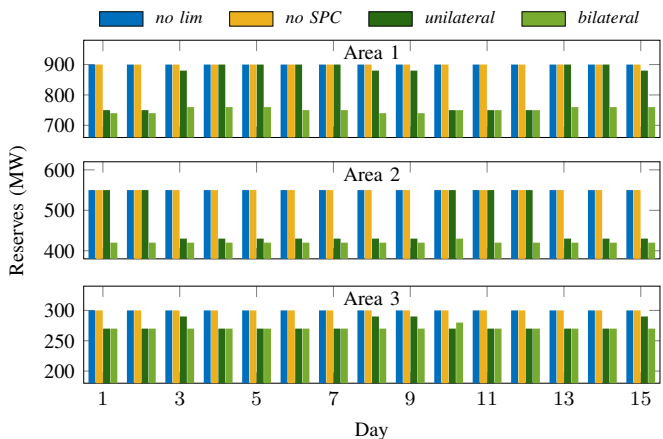
\begin{figure}[!t]
    \vspace{-0.6em}
    \resizebox{0.49\textwidth}{!}{%
        \begin{tikzpicture}
            \begin{axis}[%
                width=0.44\textwidth,
                height=0.0551\textheight,
                at={(0,0)},
                scale only axis,
                bar shift=0,
                xmin=0.5,
                xmax=15.5,
                xtick={1,3,5,7,9,11,13,15},
                xticklabels={},
                xlabel style={font=\footnotesize},
                ymin=0.66,
                ymax=0.98,
                ytick={0.7,0.8,0.9},
                yticklabels={700,800,900},
                yticklabel style={font=\footnotesize},
                ylabel style={font=\footnotesize},
                axis background/.style={fill=white},
                title style={font=\footnotesize,yshift=-0.6cm},
                title={Area 1},
                legend columns=4,
                legend style={at={(0.5,1.03)}, anchor=south, legend cell align=left, align=left, draw=none, line width=0.2, font=\scriptsize},
                /tikz/every even column/.append style={column sep=0.3cm},
                every axis legend/.append style={column sep=0.3em},
                ]
                
                \addplot[ybar, bar width=0.18, fill=blue_plot, draw=none, area legend] table[x index=0,y index=4]{Plots/Data/res3.dat};
                \addplot[ybar, bar width=0.18, fill=yellow_plot, draw=none, area legend] table[x index=1,y index=5]{Plots/Data/res3.dat};
                \addplot[ybar, bar width=0.18, fill=green1_plot, draw=none, area legend] table[x index=2,y index=6]{Plots/Data/res3.dat};
                \addplot[ybar, bar width=0.18, fill=green2_plot, draw=none, area legend] table[x index=3,y index=7]{Plots/Data/res3.dat};
                \legend{\textit{no lim},\textit{no SPC},\textit{unilateral},\textit{bilateral}}
            \end{axis}
            \begin{axis}[%
                width=0.44\textwidth,
                height=0.055\textheight,
                at={(0,-300)},
                scale only axis,
                bar shift=0,
                xmin=0.5,
                xmax=15.5,
                xtick={1,3,5,7,9,11,13,15},
                xticklabels={},
                xticklabel style={font=\footnotesize},
                xlabel style={font=\footnotesize},
                ymin=0.38,
                ymax=0.62,
                ytick={0.4,0.5,0.6},
                yticklabels={400,500,600},
                yticklabel style={font=\footnotesize},
                ylabel style={font=\footnotesize},
                ylabel={Reserves (MW)},
                axis background/.style={fill=white},
                title style={font=\footnotesize,yshift=-0.6cm},
                title={Area 2},
                ]
                \addplot[ybar, bar width=0.18, fill=blue_plot, draw=none, area legend] table[x index=0,y index=8]{Plots/Data/res3.dat};
                \addplot[ybar, bar width=0.18, fill=yellow_plot, draw=none, area legend] table[x index=1,y index=9]{Plots/Data/res3.dat};
                \addplot[ybar, bar width=0.18, fill=green1_plot, draw=none, area legend] table[x index=2,y index=10]{Plots/Data/res3.dat};
                \addplot[ybar, bar width=0.18, fill=green2_plot, draw=none, area legend] table[x index=3,y index=11]{Plots/Data/res3.dat};
            \end{axis}
            \begin{axis}[%
                width=0.44\textwidth,
                height=0.055\textheight,
                at={(0,-405)},
                scale only axis,
                bar shift=0,
                xmin=0.5,
                xmax=15.5,
                xtick={1,3,5,7,9,11,13,15},
                xticklabel style={font=\footnotesize},
                xlabel={Day},
                xlabel style={font=\footnotesize},
                ymin=0.18,
                ymax=0.34,
                ytick={0.2,0.25,0.3},
                yticklabels={200,250,300},
                yticklabel style={font=\footnotesize},
                ylabel style={font=\footnotesize},
                axis background/.style={fill=white},
                title style={font=\footnotesize,yshift=-0.6cm},
                title={Area 3},
                ]
                \addplot[ybar, bar width=0.18, fill=blue_plot, draw=none, area legend] table[x index=0,y index=12]{Plots/Data/res3.dat};
                \addplot[ybar, bar width=0.18, fill=yellow_plot, draw=none, area legend] table[x index=1,y index=13]{Plots/Data/res3.dat};
                \addplot[ybar, bar width=0.18, fill=green1_plot, draw=none, area legend] table[x index=2,y index=14]{Plots/Data/res3.dat};
                \addplot[ybar, bar width=0.18, fill=green2_plot, draw=none, area legend] table[x index=3,y index=15]{Plots/Data/res3.dat};
            \end{axis}
        \end{tikzpicture}
    }%
    \vspace{-1.2em}
    \caption{Generation capacity reserved during each day of the simulations.}
    \label{fig:reserves3}
    \vspace{-0.3em}
\end{figure}

According to the findings in \cite{SANZ2017}, we consider now that HVDC lines can be overloaded to provide frequency support (up to 20\% of the nominal rating), increasing the utilization of the interconnectors for energy exchanges. The procurement of primary reserves in the three areas is provided in \figurename~\ref{fig:reserves3}. Similar to the previous example, the exchange of reserves through HVDC results in less procurement in each area, with the results that more cheap generation is available for energy production. The total procurement costs in the case ``\textit{no SPC}'', used as reference, is 5.78 million Euros; this cost decrease to 5.22 million Euros (-9.6\%) in ``\textit{unilateral}'' and to 4.71 million Euros in ``\textit{bilateral}'' (-18.5\%). The larger decrease compared to the 2-area example can be explained with the larger capacity between area 1 and 2, and with the additional support provided by area 3. In \figurename~\ref{fig:windcorr}, the hourly profiles of wind production and HVDC support are compared: in the case of UCS, a certain correlation between high RES penetration and HVDC frequency support can be observed. This clearly shows that HVDC SPC can help maintaining the system N-1 secure in the event of low inertia periods, reducing the costs associated with RES curtailment and other security measures. 


The comparison of the two converter types, namely LCC and VSC, is provided in \figurename~\ref{fig:lccvsc}, where the flows on two interconnectors are shown for the first three days of the simulated time horizon. The additional constraints introduced for LCC converters result in unidirectional power flows for a certain time period, 24 hours in our simulation. In addition, when frequency support is provided, the flow on the interconnector must be kept above a certain threshold to avoid changes of the cable polarity. This does not happen with VSC converters, whose flows can happen in the two directions without time constraints. From \figurename~\ref{fig:lccvsc} it is also possible to see that, by considering the overloading capabilities of the lines, less capacity is reserved.

\definecolor{gray_lcc}{rgb}{0.90196,0.90196,0.90196}%
\definecolor{blue_lcc}{rgb}{0.05882,0.29412,0.45098}%
\definecolor{lightblue_lcc}{rgb}{0.5569,0.6627,0.8588}%
\definecolor{yellow_lcc}{rgb}{0.92941,0.69412,0.12549}%

\begin{figure}[!t]
    \resizebox{0.485\textwidth}{!}{%
    \begin{tikzpicture}
        
        \begin{axis}[%
            width=0.4401\textwidth,
            height=0.16\textheight,
            at={(0,0)},
            axis on top,
            xmin=1,
            xmax=72,
            xticklabels={},
            xlabel style={font=\footnotesize},
            xticklabel style={font=\footnotesize},
            yticklabel style={font=\footnotesize},
            ylabel style={font=\footnotesize,yshift=-5pt},
            ymin=-270,
            ymax=270,
            ylabel={Power (MW)},
            axis background/.style={fill=white},
            legend columns=2,
            legend style={at={(0.5,1.02)}, anchor=south, legend cell align=left, align=left, draw=none, line width=0.2, font=\scriptsize},
            /tikz/every even column/.append style={column sep=0.3cm},
            every axis legend/.append style={column sep=0.3em},
            ]
            \addplot [area legend, fill=gray_lcc, draw=gray_lcc] table[row sep=crcr]{%
                0 200 \\
                73 200 \\
                73 240 \\
                0 240 \\
            }--cycle;
            \addplot [fill=gray_lcc, draw=gray_lcc,forget plot] table[row sep=crcr]{%
                0 -200 \\
                73 -200 \\
                73 -240 \\
                0 -240 \\
            }--cycle;
            \addplot [area legend, fill=lightblue_lcc!50, draw=lightblue_lcc!50] table[x index=0,y index=1]{Plots/Data/lccvsc.dat}\closedcycle;
            \addplot [fill=white, draw=white, forget plot] table[x index=0,y index=2]{Plots/Data/lccvsc.dat}\closedcycle;
            \addplot [color=black, line width = 0.5pt, forget plot] table[row sep=crcr]{%
                0 0 \\
                73 0 \\
            };
            \addplot [color=blue_lcc, line width=1pt, line cap = round] table[x index=0,y index=3]{Plots/Data/lccvsc.dat};
            \addplot [fill=white, draw=white, forget plot] table[row sep=crcr]{%
                56 20\\
                60 20\\
            }\closedcycle;
            \addplot [color=black, line width = 0.5pt, forget plot] table[row sep=crcr]{%
                55 0 \\
                61 0 \\
            };
            \addplot [fill=gray_lcc, draw=gray_lcc,forget plot] table[row sep=crcr]{%
                23 -200 \\
                25 -200 \\
                25 -240 \\
                23 -240 \\
            }--cycle;
            \addplot [color=red, dashed, line width = 0.75pt] table[row sep=crcr]{%
                0 200 \\
                73 200 \\
            };
            \addplot [color=red, dashed, line width = 0.75pt] table[row sep=crcr]{%
                0 -200 \\
                73 -200 \\
            };
            \legend{Overload region,Reserve,Flow,Nominal rating}
        \end{axis}
        \begin{axis}[%
            width=0.44\textwidth,
            height=0.16\textheight,
            at={(0in,-1in)},
            axis on top,
            xmin=1,
            xmax=72,
            xlabel style={font=\footnotesize},
            xticklabel style={font=\footnotesize},
            yticklabel style={font=\footnotesize},
            ylabel style={font=\footnotesize,yshift=-5pt},
            xtick = {1,10,20,30,40,50,60,70},
            xticklabels = {0,10,20,30,40,50,60,70},
            xlabel={Hours (h)},
            ymin=-650,
            ymax=650,
            ylabel={Power (MW)},
            axis background/.style={fill=white},
            legend style={at={(0.424,0.936)}, anchor=south west, legend cell align=left, align=left, draw=white!15!black}
            ]
            \addplot [fill=gray_lcc, draw=gray_lcc] table[row sep=crcr]{%
                0 500 \\
                73 500 \\
                73 600 \\
                0 600 \\
            }--cycle;
            \addplot [fill=gray_lcc, draw=gray_lcc,forget plot] table[row sep=crcr]{%
                0 -500 \\
                73 -500 \\
                73 -600 \\
                0 -600 \\
            }--cycle;
            \addplot [fill=lightblue_lcc!50, draw=lightblue_lcc!50] table[x index=0,y index=4]{Plots/Data/lccvsc.dat}\closedcycle;
            \addplot [fill=white, draw=white, forget plot] table[x index=0,y index=5]{Plots/Data/lccvsc.dat}\closedcycle;
            \addplot [color=black, line width = 0.5pt, forget plot] table[row sep=crcr]{%
                0 0 \\
                73 0 \\
            };
            \addplot [color=blue_lcc, line width=1pt, line cap = round] table[x index=0,y index=6]{Plots/Data/lccvsc.dat};
            \addplot [color=red, dashed, line width = 0.75pt] table[row sep=crcr]{%
                0 500 \\
                73 500 \\
            };
            \addplot [color=red, dashed, line width = 0.75pt] table[row sep=crcr]{%
                0 -500 \\
                73 -500 \\
            };

        \end{axis}
    \end{tikzpicture}}
    \vspace{-1.6em}
    \caption{Example of flows and capacity reservation (unilateral scheme) on LCC-HVDC (upper figure) and VSC-HVDC links (lower figure).}
    \label{fig:lccvsc}
    \vspace{-1.5em}
\end{figure}
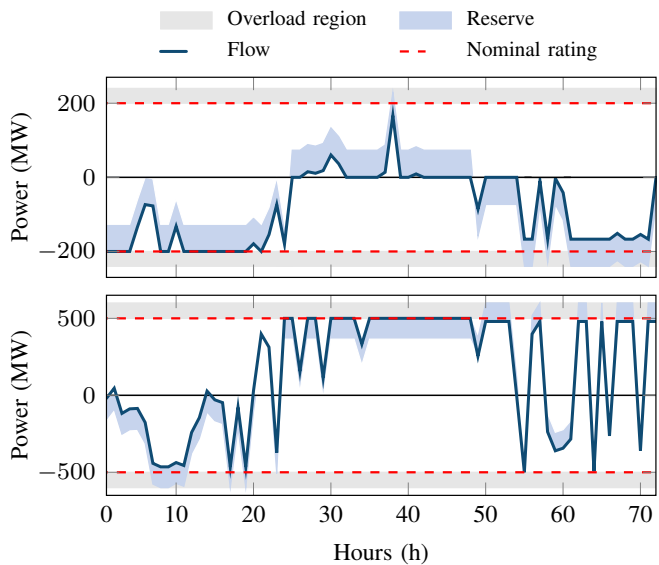

Finally, \figurename~\ref{fig:solvtime} compares the average solver time for clearing the market (24 h) with the two test cases. Clearly, introducing the SPC functionality in the formulation increases the complexity of the problem, with the result that the computation time increases. The highest solver time is seen for the ``\textit{unilateral}'' case, with an average of 26.8 minutes; however, the simulation have been run on a machine with an Intel Core 2.9 GHz CPU (4 cores, 32 GB of RAM) and better results can be expected using parallelization in high performance computers.




\section{Conclusion}\label{sec:6}
While RES are being integrated into electrical networks around the world, transmission system operators are facing serious challenges in operating the system with low inertia levels. With the ambitious climate targets set by the European Commission, coordination between system operators will be key to secure the operation of power systems during the energy transition. In this regard, the goal of this paper is to highlight the benefits of sharing reserves through HVDC interconnectors using their Supplementary Power Control (SPC) functionality. First, we have derived analytical expressions for the frequency metrics of asynchronous AC systems connected by HVDC lines. We have then introduced these metrics as constraints into a unit commitment problem. Two different control schemes for HVDC converters have been analyzed: the unilateral scheme, which allows to share reserves only in one direction, and the bilateral scheme, which allows it in both directions. The results of the optimization problem show how procurement costs can be reduced up to 20\% if the SPC functionality of HVDC lines is activated. Moreover, we have analyzed how generators and loads are impacted by these modifications, showing that the unilateral scheme might results in one-side benefits, while with a bilateral scheme the benefits are more distributed. 

\begin{figure}[!t]
    \vspace{-0.6em}
    \resizebox{0.49\textwidth}{!}{%
        \begin{tikzpicture}
            \begin{axis}[%
                width=0.44\textwidth,
                height=0.13\textheight,
                at={(0,0)},
                axis y line*=right,
                axis x line=none,
                xmin=0,
                xmax=360,
                xticklabels={},
                xlabel style={font=\footnotesize},
                ymin=0,
                ymax=5,
                yticklabel style={font=\footnotesize},
                ylabel style={font=\footnotesize},
                title style={font=\footnotesize,yshift=-0.6cm},
                ]
                \addplot [color=yellow_lcc, line width=1pt, line cap = round] table[x index=0,y index=7]{Plots/Data/windcorr.dat};
            \end{axis}
            \begin{axis}[%
                width=0.44\textwidth,
                height=0.13\textheight,
                at={(0,0)},
                axis y line*=left,
                xmin=0,
                xmax=360,
                xticklabels={},
                xlabel style={font=\footnotesize},
                ymin=0,
                ymax=220,
                yticklabel style={font=\footnotesize},
                ylabel style={font=\footnotesize},
                title style={font=\footnotesize,yshift=-0.6cm},
                title={Area 1},
                legend columns=3,
                legend style={at={(0.5,1.03)}, anchor=south, legend cell align=left, align=left, draw=none, line width=0.2, font=\scriptsize},
                /tikz/every even column/.append style={column sep=0.3cm},
                every axis legend/.append style={column sep=0.3em},
                ]
                \addplot [draw=blue_lcc, line width = 1pt] table[row sep=crcr]{%
                    -10 1 \\
                    -11 1 \\
                };
                \addplot [draw=lightblue_lcc, line width = 1pt] table[row sep=crcr]{%
                    -10 1 \\
                    -11 1 \\
                };
                \addplot [draw=yellow_lcc, line width = 1pt] table[row sep=crcr]{%
                    -10 1 \\
                    -11 1 \\
                };
                \addplot [color=lightblue_lcc, line width=1pt, line cap = round, forget plot] table[x index=0,y index=4]{Plots/Data/windcorr.dat};
                \addplot [color=blue_lcc, line width=1pt, line cap = round, forget plot] table[x index=0,y index=1]{Plots/Data/windcorr.dat};
                \legend{HVDC support (UCS),HVDC support (BCS),Wind production}
            \end{axis}
            
            \begin{axis}[%
                width=0.44\textwidth,
                height=0.13\textheight,
                at={(0in,-0.8in)},
                axis y line*=right,
                axis x line=none,
                xmin=0,
                xmax=360,
                xticklabels={},
                xlabel style={font=\footnotesize},
                ymin=0,
                ymax=2.5,
                yticklabel style={font=\footnotesize},
                ylabel style={font=\footnotesize,yshift=-3pt},
                ylabel={Wind production (GW)},
                title style={font=\footnotesize,yshift=-0.6cm},
                ]
                \addplot [color=yellow_lcc, line width=1pt, line cap = round] table[x index=0,y index=8]{Plots/Data/windcorr.dat};
            \end{axis}
            \begin{axis}[%
                width=0.44\textwidth,
                height=0.13\textheight,
                at={(0in,-0.8in)},
                axis y line*=left,
                xmin=0,
                xmax=360,
                xticklabels={},
                xlabel style={font=\footnotesize},
                ymin=0,
                ymax=220,
                yticklabel style={font=\footnotesize},
                ylabel style={font=\footnotesize},
                ylabel={HVDC support (MW)},
                title style={font=\footnotesize,yshift=-0.6cm},
                title={Area 2},
                ]
                \addplot [color=lightblue_lcc, line width=1pt, line cap = round, forget plot] table[x index=0,y index=5]{Plots/Data/windcorr.dat};
                \addplot [color=blue_lcc, line width=1pt, line cap = round, forget plot] table[x index=0,y index=2]{Plots/Data/windcorr.dat};
            \end{axis}
            
            \begin{axis}[%
                width=0.44\textwidth,
                height=0.13\textheight,
                at={(0in,-1.6in)},
                axis y line*=right,
                axis x line=none,
                xmin=0,
                xmax=360,
                xlabel style={font=\footnotesize},
                ymin=0,
                ymax=1,
                yticklabel style={font=\footnotesize},
                ylabel style={font=\footnotesize},
                title style={font=\footnotesize,yshift=-0.6cm},
                ]
                \addplot [color=yellow_lcc, line width=1pt, line cap = round] table[x index=0,y index=9]{Plots/Data/windcorr.dat};
            \end{axis}
            \begin{axis}[%
                width=0.44\textwidth,
                height=0.13\textheight,
                at={(0in,-1.6in)},
                axis y line*=left,
                xmin=0,
                xmax=360,
                xticklabels={},
                xlabel style={font=\footnotesize},
                ymin=0,
                ymax=50,
                yticklabel style={font=\footnotesize},
                ylabel style={font=\footnotesize},
                title style={font=\footnotesize,yshift=-0.6cm},
                title={Area 3},
                ]
                \addplot [color=lightblue_lcc, line width=1pt, line cap = round, forget plot] table[x index=0,y index=6]{Plots/Data/windcorr.dat};
                \addplot [color=blue_lcc, line width=1pt, line cap = round, forget plot] table[x index=0,y index=3]{Plots/Data/windcorr.dat};
            \end{axis}
            
        \end{tikzpicture}
    }%
    \vspace{-1.2em}
    \caption{Correlation between HVDC support and wind power production.}
    \label{fig:windcorr}
\end{figure}
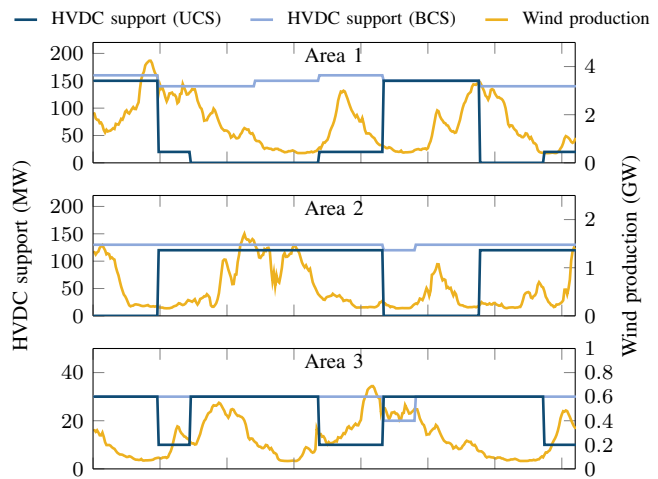
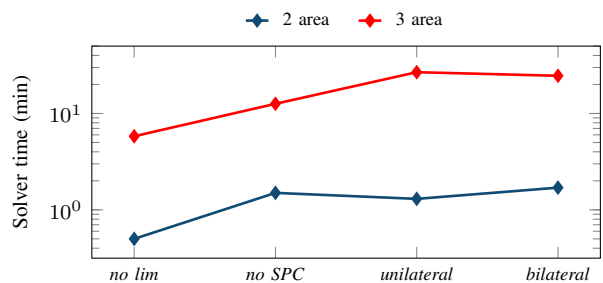
\begin{figure}[!t]
        \begin{tikzpicture}
            \begin{axis}[%
                width=0.46\textwidth,
                height=0.18\textheight,
                at={(0,0)},
                xmin=0.7,
                xmax=4.3,
                xtick={1,2,3,4},
                xticklabels={\textit{no lim},\textit{no SPC},\textit{unilateral},\textit{bilateral}},
                xticklabel style={font=\scriptsize},
                xlabel style={font=\footnotesize},
                ymin=0,
                ymax=50,
                ymode=log,
                yticklabel style={font=\footnotesize},
                ylabel style={font=\footnotesize},
                ylabel={Solver time (min)},
                legend columns=2,
                legend style={at={(0.5,1.03)}, anchor=south, legend cell align=left, align=left, draw=none, line width=0.2, font=\scriptsize},
                /tikz/every even column/.append style={column sep=0.3cm},
                every axis legend/.append style={column sep=0.3em},
                ]
                \addplot [color=blue_lcc, line width = 1pt, mark=diamond*] table[row sep=crcr]{%
                    1 0.5 \\
                    2 1.5 \\
                    3 1.3 \\
                    4 1.7 \\
                };
                \addplot [color=red, line width = 1pt, mark=diamond*] table[row sep=crcr]{%
                    1 5.8 \\
                    2 12.6 \\
                    3 26.8 \\
                    4 24.6 \\
                };
                \legend{2 area, 3 area}
            \end{axis}
            
        \end{tikzpicture}
    \vspace{-0.6em}
    \caption{Solver time comparison.}
    \label{fig:solvtime}
    \vspace{-0.3em}
\end{figure}

As future research, the proposed model could be expanded in two directions. First, by also including hydropower plants, which will allow to consider the valve backslash effect of hydro units in the system frequency response. Hydro generators have, so far, not been considered in any of the existing approaches combining unit commitment and frequency response in the literature. Second, the inclusion of multi-terminal DC grids could represent the possibility of simultaneously sharing reserves among several asynchronous regions.



\appendices
\section{Calculation of the hyperplane coefficients}\label{appendix:A}

In this section, the different steps for the calculation of the hyperplane coefficients are explained. The calculation is done for the unilateral scheme, using Eq. \eqref{eq:ifd_unil_a} for the calculation of the IFD; however, the same procedure applies for the bilateral scheme.  

First, the three system parameters appearing in Eq. \eqref{eq:ifd_unil_a}, $\widehat{R}_a$, $\widehat{F}_a$ and $\widehat{M}_a$, are discretized. The range used for the discretization depends on the parameters of the generators in the area, and the number of points per parameter on the desired accuracy, e.g., with 100 points per parameter, the IFD function is evaluated in 1 million points. The IFD is calculated with each triplet of parameters, and the spatial representation of the IFD function is obtained. 

The second step consists in the selection of the points which are close to the IFD limits, which constitute the set $\mathcal{S}$. For example, if the maximum allowed IFD is 0.7 Hz, all the points in the range $0.7\pm0.01$ Hz are selected. These points are used for the calculation of the hyperplane that better interpolates them. This is done with the following optimization problem:

\vspace{-0.6pt}{\small
\begin{subequations}
\begin{alignat}{2}
    & \underset{A^{\textsc{R}}_a,A^{\textsc{M}}_a,A^0_a}{\text{min}} \enspace && \sum_{(i,j,k)\in\mathcal{S}} (A^{\textsc{R}}_a \widehat{R}_{i,a} + A^{\textsc{M}}_a \widehat{M}_{j,a} + A^0_a - \widehat{F}_{k,a})^2 \label{apA:obj}\\
    & \text{s.t.} \enspace \enspace && A^{\textsc{R}}_a \widehat{R}_{i,a} + A^{\textsc{M}}_a \widehat{M}_{j,a} + A^0_a - \widehat{F}_{k,a} \leq 0 \label{apA:ineq}
\end{alignat}%
\end{subequations}%
}%
where the objective is to find the coefficients $A^{\textsc{R}}_a$, $A^{\textsc{M}}_a$ and $A^0_a$ which minimize the distance between the hyperplane and the points in $\mathcal{S}$. Constraint \eqref{apA:ineq} guarantees that the hyperplane lies completely in the feasible space, i.e. the space identified by the points with IFD$\leq$0.7 Hz. In this way, some points that would satisfy the IFD constraints might be excluded but, most importantly, no triplet ($\widehat{R}_a$,$\widehat{F}_a$,$\widehat{M}_a$) which violates the IFD constraints is included in the feasible space of the problem. 
\bibliographystyle{myIEEEtran.bst}
\bibliography{00_main.bib}{}

\end{document}